\documentclass[12pt]{article}

\usepackage{newtxtext,newtxmath}

\usepackage{graphicx}

\usepackage[letterpaper,margin=1in]{geometry}

\renewenvironment{abstract}
	{\quotation}
	{\endquotation}

\date{}

\makeatletter
\renewcommand{\fnum@figure}{\textbf{Figure \thefigure}}
\renewcommand{\fnum@table}{\textbf{Table \thetable}}
\makeatother

\usepackage{scicite}

\usepackage{url}

\newcommand{\bm}[1]{\mbox{\boldmath $#1$}}	

\def\scititle{
    Frustrated junctions in interfacial networks
}
\title{\bfseries \boldmath \scititle}

\author{
	H{\aa}kan~Hallberg$^{1\ast}$,
	Vasily~V.~Bulatov$^{2}$,
	Bryan~W.~Reed$^{3}$,
	Mukul~Kumar$^{2}$\and
	\small$^{1}$Division of Solid Mechanics, Lund University, Lund, Sweden.\and\
	\small$^{2}$Lawrence Livermore National Laboratory, Livermore, USA.\and
	\small$^{3}$Integrated Dynamic Electron Solutions, Inc., Pleasanton, USA.\and
	\small$^\ast$Corresponding author. Email: hakan.hallberg@solid.lth.se
}

\begin{document} 

\maketitle

\begin{abstract} \bfseries \boldmath
Networks of interfaces in materials and living systems evolve through motion and rearrangement of interfaces and junctions. Local equilibrium at a junction requires interfacial force balance. We show that in three dimensions the classical Herring condition for triple lines is necessary but insufficient: four triple lines meeting at a quadruple node may each satisfy Herring equilibrium, while the four conditions remain mutually incompatible. Such a frustrated node has no admissible local equilibrium geometry. In the semi-isotropic limit, the six interfacial energies must form the edge lengths of a non-degenerate tetrahedron, with constructibility determined by the Cayley--Menger determinant. This hidden compatibility constraint arises from three-dimensional geometry and applies broadly to foams, polycrystals, tissues and other interfacial networks.
\end{abstract}

\newpage

\noindent
Three-dimensional interfacial networks partition space into cells, such as bubbles in liquid foams, grains in man-made polycrystalline materials or natural minerals, and cells in biological tissues \cite{Weaire1984,Glazier1992,Farhadifar2007,Stavans1993,Weaire2015}. One-dimensional lines where three interfaces meet are often called triple lines but, since lines where more than three interfaces meet are also occasionally observed, we will refer to such one-dimensional objects as $n$-lines. Similarly, four 3-lines meet at zero-dimensional quadruple nodes, but $n$-nodes connecting more than four lines are also occasionally observed. Typically unstable, $n$-lines with $n>3$ and $n$-nodes with $n>4$ occur in transients when the network locally rearranges its topology. Together, lines and nodes where interfaces meet are commonly called interfacial junctions.

\begin{figure}
	\centering
	\includegraphics[width=\textwidth]{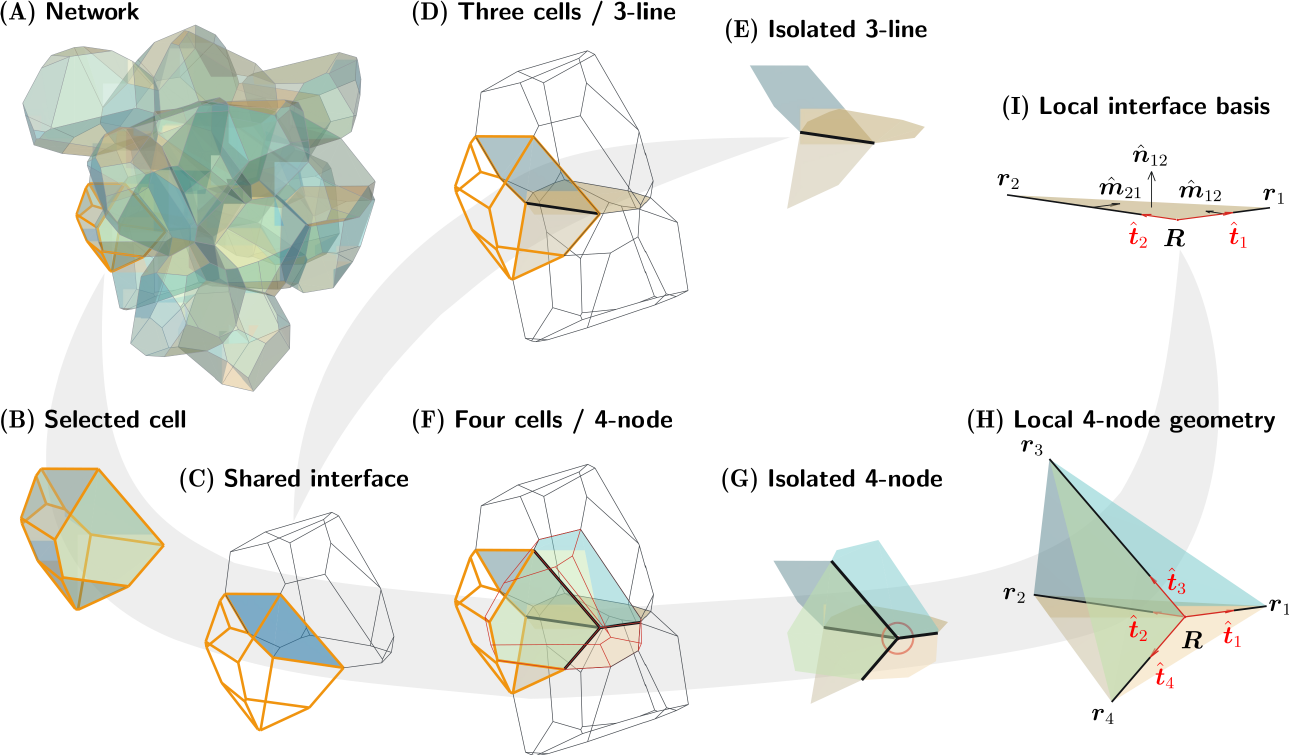}
	\caption{\textbf{Hierarchy of cellular-network elements and local 4-node notation.} (\textbf{A}) Representative three-dimensional cellular network. Three-dimensional cells are separated by two-dimensional interfaces; intersections of interfaces form one-dimensional lines, and intersections of lines form zero-dimensional nodes. (\textbf{B}) Single-cell cutout from the network in (\textbf{A}), with the highlighted skeleton showing the lines and nodes incident to the cell. (\textbf{C}) Two neighboring cells sharing a common interface, the elementary two-dimensional boundary between adjacent cells. (\textbf{D})--(\textbf{E}) Three cells incident to a common 3-line. The isolated view in (\textbf{E}) shows the 3-line as the intersection of three interfaces. (\textbf{F})--(\textbf{G}) Four cells incident to a common 4-node. The isolated view in (\textbf{G}) shows the local 4-node geometry: six pairwise interfaces, four 3-lines and one 4-node, circled in red. (\textbf{H}) Local 4-node geometry used in the derivation. The 4-node is positioned at $\bm{R}$ and connects four 3-lines. A small control volume around the 4-node is spanned by vectors $\bm{r}_{i}=l_{i}\hat{\bm{t}}_{i}$, $i=1,2,3,4$, where $\hat{\bm{t}}_{i}$ are unit tangents directed away from the 4-node. The six interface patches are locally planar and have unit normals $\hat{\bm{n}}_{ij}=(\hat{\bm{t}}_{i}\times\hat{\bm{t}}_{j})/|\hat{\bm{t}}_{i}\times\hat{\bm{t}}_{j}|$. (\textbf{I}) Local notation for one interface patch, spanned by $\hat{\bm{t}}_1$ and $\hat{\bm{t}}_2$, with unit normal $\hat{\bm{n}}_{12}$ pointing out of the plane. The in-plane vector $\hat{\bm{m}}_{12}=\hat{\bm{n}}_{12}\times\hat{\bm{t}}_1$ is perpendicular to 3-line 1 and gives the capillary-force direction exerted by interface 12 on 3-line 1. The corresponding vector on 3-line 2 is $\hat{\bm{m}}_{21}=\hat{\bm{n}}_{21}\times\hat{\bm{t}}_2$, with $\hat{\bm{n}}_{21}=-\hat{\bm{n}}_{12}$.}
	\label{fig_001}
\end{figure}

In isotropic cellular networks such as soap foams, the geometry of interfaces, lines and nodes is described by Plateau’s laws of minimal surfaces \cite{Taylor1976}. 4-nodes also arise naturally in multiphase systems \cite{Gui2008}. In polycrystalline networks, the corresponding quadruple-point topology\textemdash four grains, six grain boundaries and four triple lines\textemdash was described explicitly by Rhines \cite{Rhines1970}, and has more recently been resolved directly in three-dimensional grain-boundary reconstructions \cite{CorderoBorboa2024}. Fig.~\ref{fig_001}A--G illustrates the hierarchy of topological elements in a cellular network, from cells to interfaces to 3-lines and 4-nodes \cite{Lazar2012}.

\subsection*{Local 4-node equilibrium}
Following Herring \cite{Herring1951,herring1951some} and the Cahn--Hoffman capillarity-vector formulation for anisotropic interfaces \cite{Hoffman1972}, the equilibrium for an individual 3-line is well understood: its three interfacial tensions, supplemented by inclination-torque terms when the interfacial energies depend on plane orientation, should close as a force triangle \cite{Marks2012}. Rather than re-deriving Herring's classical equation, we will now show that it emerges as a necessary condition for the equilibrium of a 4-node. In the derivation below, $i,j=1,2,3,4$ label the four 3-lines emanating from the 4-node, and the interface $ij$ is bounded by 3-lines $i$ and $j$. Fig.~\ref{fig_001}H shows the idealized local 4-node control volume, with each 3-line terminating at $\bm{r}_{i}$. Fig.~\ref{fig_001}I defines the local basis associated with one representative interface patch. The energy associated with the 4-node control volume is 
\begin{equation}
	E = \sum_{i=1}^3\sum_{j = i+1}^4 A_{ij}\gamma_{ij}(\hat{\bm{n}}_{ij}),
	\label{eqn_001}
\end{equation}
where $A_{ij}$ is the area of the triangle formed by the central 4-node and two end points $\bm{r}_i$ and $\bm{r}_j$, $\gamma_{ij}$ is the energy per unit area of the interface $ij$ that is bounded by 3-lines $i$ and $j$ and the sum is over all six interfaces that meet at the 4-node. The length of the $i$-th 3-line is $l_{i}=|\bm{r}_{i}|$. With the vector-calculus details given in \cite{methods}, the force associated with displacing the 4-node is the sum of six interface contributions, e.g., for interface $12$: 
\begin{equation}
	\bm{f}_{12}
	=
	l_1\left[\gamma_{12}\hat{\bm{m}}_{12} + \hat{\bm{n}}_{12}	\left(\hat{\bm{m}}_{12}\cdot\nabla_{\hat{\bm{n}}}\gamma_{12}\right)\right]
	+
	l_2\left[
	\gamma_{12}\hat{\bm{m}}_{21} + \hat{\bm{n}}_{21}\left(\hat{\bm{m}}_{21}\cdot\nabla_{\hat{\bm{n}}}\gamma_{12}\right)\right],
	\label{eqn_002}
\end{equation}
and similarly for interfaces 13, 14, 23, 24 and 34. Each interface contribution contains two terms, each associated with one of two 3-lines bounding the interface. The resulting 12 terms can now be re-grouped around four 3-lines, with interfacial contributions associated with each 3-line, as follows
\begin{equation}
	\bm{f}_{\bm{R}}	= \sum_{1\le i<j\le4}\bm{f}_{ij} = \sum_{i=1}^{4}l_i\bm{H}_{i},
	\qquad
	\bm{H}_{i} = \sum_{j\ne i} \left[\gamma_{ij}\hat{\bm{m}}_{ij} +
	\hat{\bm{n}}_{ij}\left(\hat{\bm{m}}_{ij}\cdot\nabla_{\hat{\bm{n}}}\gamma_{ij}\right)\right].
	\label{eqn_003}
\end{equation}
Each Herring residual vector $\bm{H}_{i}$ is the sum of contributions from three interfaces meeting at 3-line $i$ and is the net capillary force per unit length on the same 3-line in the adopted sign convention. The only difference between $\bm{H}_{i}$ as written above and the equation familiar from classical works and textbooks \cite{herring1951some,porter2021phase,King1999} is in notation: here, we enumerate four 3-lines meeting at a 4-node instead of enumerating three interfaces meeting at a 3-line, as in Herring's derivation.  

For a given finite 4-node control volume (see Fig.~\ref{fig_001}H), the force conjugate to a displacement of the 4-node vanishes when the length-weighted sum in Eq.~(\ref{eqn_003}) is zero. However, to be local, such an equilibrium should depend only on the geometry of an infinitesimal neighborhood around the node, but not on the arbitrary lengths $l_i$ of the four line segments. Therefore, scale-invariant local equilibrium requires all four vectors $\bm{H}_i=\bm{0}$, $i=1,2,3,4$. Although these conditions have previously been derived from related considerations \cite{Mason2017}, it has been tacitly assumed that the four equations can always be satisfied simultaneously by one common local geometry. In what follows, we show that such an assumption is incorrect and derive conditions under which no equilibrium can be achieved at a 4-node regardless of its local geometry. More specifically, we show that, when considered in isolation, each of the four 3-lines meeting at a 4-node can satisfy its individual force balance condition $\bm{H}_i = \bm{0}$, while no three-dimensional equilibrium geometry at the same 4-node may exist in which the forces on the four 3-lines are \emph{all simultaneously zero}. We refer to such incompatible 4-nodes as frustrated junctions. Node incompatibility described here is distinct from crystallographic $\Sigma$ combination/exclusion rules in grain-boundary networks in crystals of cubic symmetry \cite{Reed2004}.  

\subsection*{Tetrahedral constructibility in the semi-isotropic limit}
\begin{figure}
	\centering
	\includegraphics[width=\textwidth]{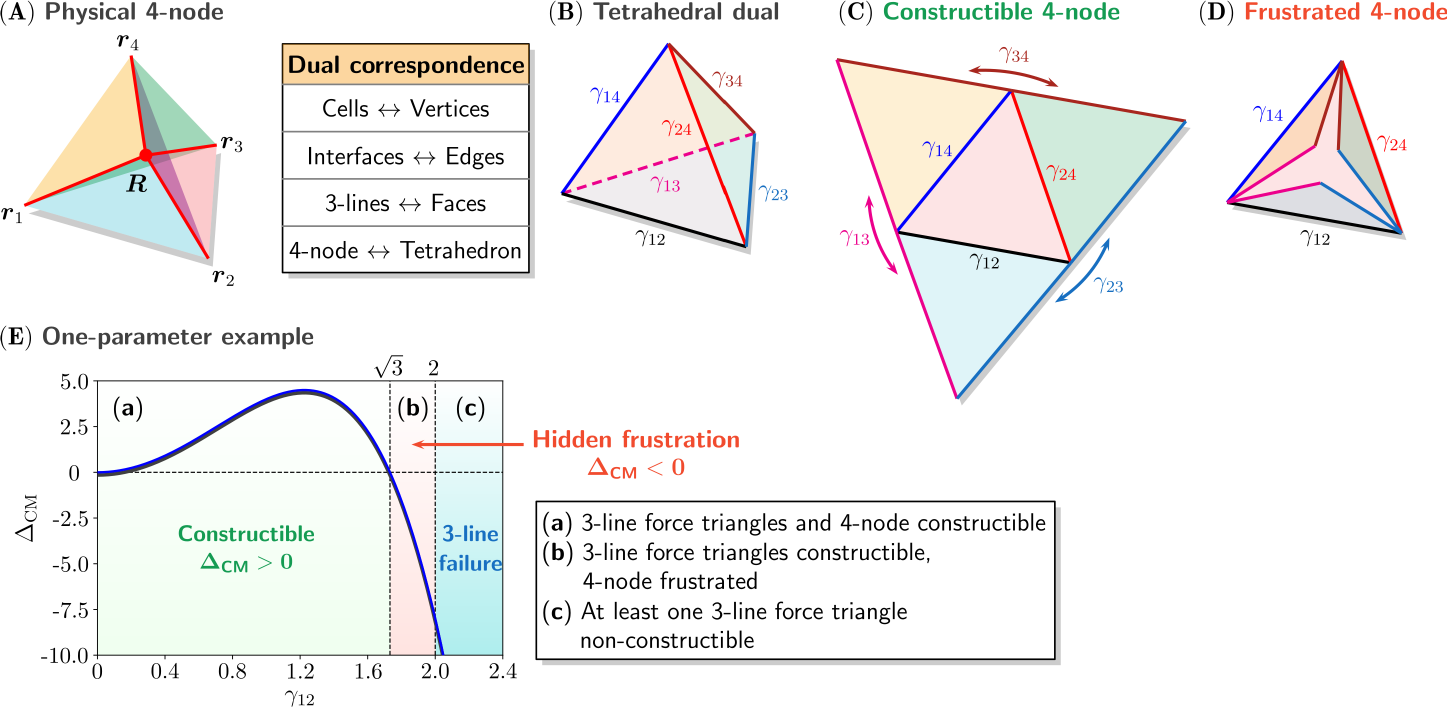}
	\caption{\textbf{Local equilibrium of all 3-lines is necessary but not sufficient for a 4-node.} (\textbf{A}) A 4-node with 3-lines shown in red. (\textbf{B}) Tetrahedral dual of the 4-node, with edge lengths corresponding to interface energies. (\textbf{C}) Foldable cutout for a constructible 4-node. (\textbf{D}) Non-foldable cutout: the triangular faces remain individually constructible, but the full tetrahedral dual is not. (\textbf{E}) One-parameter example in which $\gamma_{12}$ varies while all other interface energies equal one. Region (\textbf{b}), $\sqrt{3}<\gamma_{12}<2$, shows hidden non-constructibility: all triangular faces are individually constructible, but the 4-node is not.}
	\label{fig_002}
\end{figure}

4-node incompatibility conditions become especially transparent when the interfacial energies are independent of interface inclination. The torque terms in Eq.~(\ref{eqn_003}) then vanish, and each equation reduces to
\begin{equation}
    \bm{H}_i = \sum_{j\neq i}\gamma_{ij} \hat{\bm{m}}_{ij} = \bm{0}, \quad i=1,2,3,4.
    \label{eqn_004}
\end{equation}
The above force-balance conditions can be represented using the classical duality between a vertex shared by four three-dimensional cells and a tetrahedron whose four vertices represent those cells, as in the Delaunay--Vorono{\"i} construction \cite{Delaunay1934}. Under this duality (see Fig.~\ref{fig_002}A--B), the four cells meeting at the 4-node correspond to the four tetrahedral vertices, the six pairwise interfaces to the six edges, the four 3-lines to the four triangular faces, and the 4-node itself to the tetrahedron. Here we additionally assign to each dual edge $ij$ a length equal to the corresponding interfacial energy $\gamma_{ij}$. The four capillary forces can vanish simultaneously if and only if the six energies $\gamma_{ij}$ can form the edge lengths of a non-degenerate (foldable) tetrahedron. In this semi-isotropic setting, these descriptions are equivalent: compatibility refers to simultaneous satisfaction of the four Herring conditions, constructibility to existence of the corresponding tetrahedral dual and foldability to closure of its planar cutout into a non-degenerate tetrahedron. This is the standard edge-length criterion for tetrahedra: the four face triples must be triangular and the corresponding Cayley--Menger determinant must be positive \cite{Wirth2009}. The squared volume $\Omega_4$ of any such tetrahedron is determined by the Cayley--Menger determinant,
\begin{equation}
	288\Omega_{4}^{2} =
	\Delta_{\text{CM}} = 
	\left|
	\begin{array}{ccccc}
		0 & 1               & 1               & 1               & 1               \\
		1 & 0               & \gamma^{2}_{12} & \gamma^{2}_{13} & \gamma^{2}_{14} \\  
		1 & \gamma^{2}_{12} & 0               & \gamma^{2}_{23} & \gamma^{2}_{24} \\
		1 & \gamma^{2}_{13} & \gamma^{2}_{23} & 0               & \gamma^{2}_{34} \\
		1 & \gamma^{2}_{14} & \gamma^{2}_{24} & \gamma^{2}_{34} & 0               \\
	\end{array}\right| .
	\label{eqn_005}
\end{equation}
For a constructible tetrahedron $\Delta_{\text{CM}}>0$, $\Delta_{\text{CM}}=0$ corresponds to a limit of zero-volume ``flat'' tetrahedron and $\Delta_{\text{CM}}<0$ indicates non-constructibility. Equivalently, a positive $\Delta_{\text{CM}}$ means that the 4-node can attain local equilibrium, whereas a negative $\Delta_{\text{CM}}$ means that for such a 4-node no local equilibrium geometry exists (see Fig.~\ref{fig_002}A--D).  The frustration is hidden by the fact that all four triangular faces are individually constructible and yet the negative $\Delta_{\text{CM}}$ signals the incompatibility of the 4-node as a whole. This additional constraint has no analogue for a single triple line where triangle inequalities are both necessary and sufficient. At a 4-node, the same triangle inequalities remain necessary but do not guarantee mutual compatibility of the zero force conditions on the four 3-lines. 

The distinction is illustrated in a simple example in which five of the six interfacial energies are set to unity and only $\gamma_{12}$ is varied, in which case
\begin{equation}
	\Delta_{\text{CM}} = 2\gamma_{12}^{2}\left(3-\gamma_{12}^{2}\right).
	\label{eqn_006}
\end{equation}
In the interval $\sqrt{3} < \gamma_{12} < 2$ all four 3-line force triangles remain individually constructible or admissible, but $\Delta_{\text{CM}}$ is negative, meaning that the 4-node has no admissible equilibrium geometry, see Fig.~\ref{fig_002}E. This example motivates the statistical sampling of energy space discussed below.

\subsection*{Geometric modes and severity of frustration}
The geometric analogy with a tetrahedron can be further exploited to better understand why and how a 4-node can be frustrated. One useful classification is obtained by examining three ``corner'' angles that meet at a vertex of a flat tetrahedral cutout. Because each angle equals the geodesic length of the corresponding arc on the unit sphere centered on the vertex, we refer to the following as triangle inequalities: at each vertex, the geodesic lengths of the three spherical-triangle sides must satisfy the ordinary triangle-length inequalities. Thus, none of the three angles, or equivalently no one arc, may exceed the sum of the other two for a proper trihedral angle at the vertex to exist. It turns out that, when all four faces of a tetrahedral cutout are individually constructible (each forming a proper triangle), such triangle inequalities can be either satisfied at all four vertices (foldable cutout), fail at all four vertices at once (F4 failure), or hold at one and fail at three vertices (F3 failure).  Complete geometric conditions for cutout foldability, including a proof that no other combinations of vertex failures are possible, are given in the Supplementary Materials \cite{methods}. We also show that in an F3 cutout, the sole vertex satisfying the angle inequalities lies opposite face $i$, whose area exceeds the combined areas of the other three faces,
\begin{equation}
	\mathcal{A}_{i} > \sum_{j\ne i}\mathcal{A}_{j} .
	\label{eqn_010}
\end{equation}
When one face of the tetrahedral dual is oversized relative to the other three, the associated 3-line is defined by a comparatively high-energy combination of three interfaces. Such a 3-line is therefore a natural candidate for a wetting-like rearrangement, resulting in its removal or replacement. The specific pathway is likely to depend on the surrounding geometry and kinetics, but is analogous to replacing a high-energy (wetting) interface by two interfaces with a lower combined energy.

F4 failures have a distinct structure. At each vertex, the failed triangle-type inequality identifies one tetrahedral edge opposite the largest corner angle. In F4 cases, the four vertex failures identify two such edges, each appearing twice. Two offending edges are never in the same triangle but opposite each other in the cutout. Thus, F4 frustration implicates a pair of opposite, relatively high-energy interfaces, but does not determine a unique kinetic pathway for resolving the frustration.

It is of interest to quantify the strength of frustration, e.g., using the following normalized measure
\begin{equation}
	\chi = -\frac{\Delta_{\text{CM}}}{\langle \gamma \rangle^6}~,
	\label{eqn_007}
\end{equation}
in which the mean interfacial energy is defined as
\begin{equation}
	\langle \gamma \rangle = \frac{1}{6}\sum_{i<j}\gamma_{ij}~.
	\label{eqn_008}
\end{equation}
The quantity $\chi$ is positive for a frustrated node and is independent of the overall energy scale.

A second scale-invariant measure of frustration is the distance to the nearest constructible tetrahedron. Let $\bm{\gamma}=(\gamma_{12},\gamma_{13},\gamma_{14},\gamma_{23},\gamma_{24},\gamma_{34})$ denote the frustrated energy sextuplet and let $\bm{\gamma}'$ denote a trial sextuplet on the constructible boundary. We define
\begin{equation}
    d_{\log}(\bm{\gamma}) = \min_{\bm{\gamma}'}\left[\sum_{i<j}\left(\log\frac{\gamma'_{ij}}{\gamma_{ij}}\right)^2\right]^{1/2}, \qquad \Delta_{\rm CM}(\bm{\gamma}')=0 .
	\label{eqn_009}
\end{equation}
The logarithmic form measures the smallest relative change in the six interfacial energies required to bring the frustrated cutout $\bm{\gamma}$ to the constructible boundary. In this minimization, the trial cutouts $\bm{\gamma}'$ are assumed to have all their face-triangle inequalities satisfied and to lie on a hypersurface separating constructible $\bm{\gamma}$-sextuplets from non-constructible ones. Alternative measures of nodal frustration are described in the Supplementary Text.

\begin{figure}
	\centering
    \includegraphics[width=\textwidth]{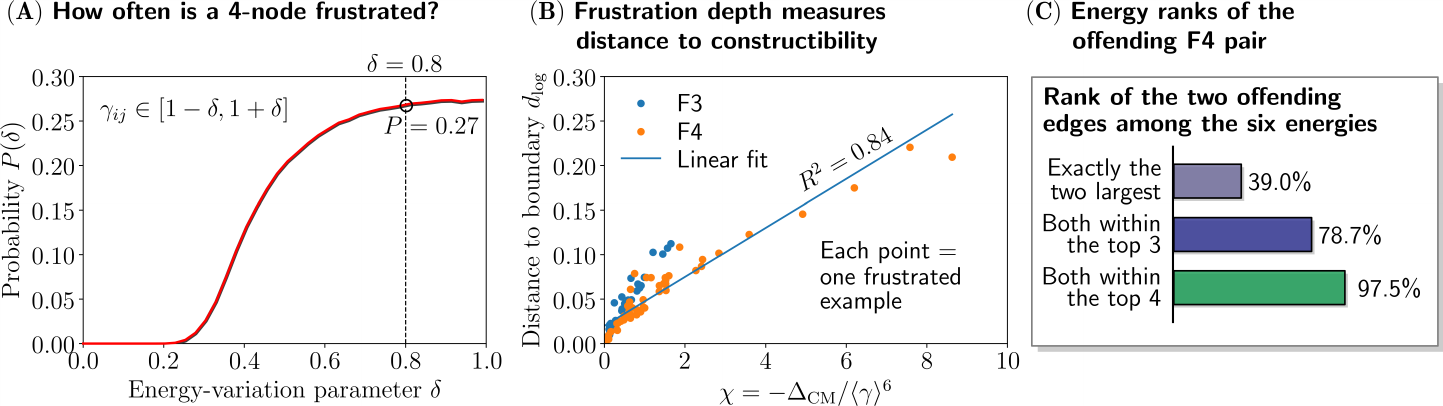}
	\caption{\textbf{Statistical occurrence, severity and structure of frustrated 4-nodes.} (\textbf{A}) Conditional probability $P(\delta)$ that a 4-node is frustrated given that all four constituent 3-line force triangles are constructible, for random six-energy sets with $\gamma_{ij}\in[1-\delta,1+\delta]$.
    At $\delta=0.8$, $P(0.8)\approx0.27$ and among samples with four constructible 3-line force triangles, the constructible, F3 and F4 fractions are 73.0\%, 2.7\% and 24.3\%, respectively.
    (\textbf{B}) Distance $d_{\log}$ to the constructible boundary versus normalized Cayley--Menger depth $\chi=-\Delta_{\rm CM}/\langle\gamma\rangle^6$ for representative frustrated samples. The linear fit gives $R^2=0.84$.
    (\textbf{C}) Rank statistics of the offending opposite-edge pair in F4 cases. The pair consists of the two largest energies in 39.0\% of cases, while both offending energies lie among the three largest in 78.7\% and among the four largest in 97.5\% of cases.}
	\label{fig_003}
\end{figure}

A natural question is how frequently frustrated 4-nodes occur. Since the constructibility conditions are homogeneous in the six interfacial energies, the relevant quantity is the relative variation among the six interfacial energies assigned to the tetrahedral dual of a node. Random sampling of $\gamma_{ij}\in[1-\delta,1+\delta]$ shows that hidden frustration becomes increasingly probable as the energy-space half-width $\delta$ increases (Fig.~\ref{fig_003}A). Here, $P(\delta)$ denotes the conditional probability that a 4-node is frustrated given that all four constituent 3-line force triangles are constructible. At $\delta=0.8$, $2\times 10 ^{6}$ sampled energy sextuplets yielded 554,951 cases with four constructible force triangles; among these, 73.0\% were constructible 4-nodes, 2.7\% were frustrated F3 nodes and 24.3\% were frustrated F4 nodes, giving a total conditional frustration probability of approximately 27.0\%. The normalized depth $\chi$ correlates strongly with the distance $d_{\log}$ to the constructible boundary (Fig.~\ref{fig_003}B), showing that $\chi$ is not only a yes-or-no indicator, but also a quantitative measure of nodal frustration. The offending pair comprised the two largest interface energies in only 39\% of F4 frustrations, but both of their energies were among the four largest in approximately 97.5\% of F4 sextuplets (Fig.~\ref{fig_003}C).

Junction frustration is favored by broad interfacial energy distributions and suppressed in nearly isotropic networks. Cellular networks such as ideal dry foams, whose local structure is governed by surface-tension-driven interface geometry, provide the isotropic limiting case: when all interfacial energies are equal, the six equal energies form a regular tetrahedral dual and the 4-node compatibility condition is automatically satisfied \cite{Matzke1945,Taylor1976,Hilgenfeldt2001}. The situation with liquid emulsions or tissue networks is less clear and deserves further investigation. In crystalline materials, however, 3-lines and 4-nodes have long been recognized as distinct geometric and thermodynamic elements of the grain-boundary network \cite{Rhines1970,King1999}. In these networks, large variations in interfacial energy are common \cite{wolf1991structure,Bulatov2014}, 
making such anisotropic networks more prone to nodal frustration \cite{Rohrer2011,Hallberg2024b}. The propensity to nodal frustration in anisotropic networks is defined not by the absolute interfacial energy scale but by the width and shape of the normalized energy distribution, sampled by interfaces present in the network \cite{rohrer2007distribution}. 

Nodal frustration is not restricted to the semi-isotropic limit discussed in some detail above. In the fully anisotropic case, the tension terms $\gamma_{ij}\hat{\bm{m}}_{ij}$ in Eq. (\ref{eqn_004}) are replaced by the full capillarity contributions in Eq. (\ref{eqn_003}), including inclination-torque terms. A fully anisotropic 4-node can attain local equilibrium if there exists a non-degenerate set of four 3-line directions for which all four anisotropic Herring residuals vanish. If no such geometry exists, the node is frustrated. A graphical extension of the dual construction and a corresponding numerical formulation of this anisotropic existence problem are provided in the Supplementary Text. Unlike the semi-isotropic case, however, this existence problem does not reduce to a simple analytical criterion such as $\Delta_{\text{CM}}>0$, and generally has to be tested numerically. Thus, the semi-isotropic result should be viewed as an analytically solvable limit of a broader existence problem.

\subsection*{Conclusions}
A constructible 4-node can, in principle, attain local equilibrium through ordinary motion of interfaces and 3-lines, whereas a frustrated 4-node cannot do so without changing its geometry, topology, interfacial energies or surrounding network. It may therefore move, develop a wetting-like configuration, collapse a local patch, or undergo a topological rearrangement of the type familiar from grain-boundary network evolution \cite{Rhines1970}. This is consistent with previous thermodynamic analyses of quadruple-junction dissociation in two-dimensional polycrystals \cite{Fortes1993}. The constructibility conditions just discussed do not always determine the pathway by which the frustration can resolve itself, but identify where ordinary relaxation is impossible. This distinction may be important in grain growth and in the more general dynamics of coarsening cellular networks \cite{Glazier1992,Macpherson2007}. Evolving cellular networks contain many ordinary perturbations in which interfaces and junctions deviate from attainable equilibria and gradually relax. Frustrated junctions are different: they are not disturbed equilibria, but local configurations for which equilibrium is geometrically unavailable. The central result is a previously unrecognized compatibility principle for three-dimensional interfacial networks: the existence of local equilibrium for every constituent 3-line does not guarantee that equilibrium of the 4-node as a whole is attainable. In the semi-isotropic limit, the missing condition is the tetrahedral constructibility of the six interfacial energies. This hidden constraint is a direct consequence of three-dimensional geometry and provides a natural framework for identifying and quantifying frustrated junctions in foams, polycrystals, tissues and other cellular systems.




\clearpage

\bibliography{science_template}

@Misc{methods,
  note = {Materials and methods are available as supplementary material},
}

@ARTICLE{Bulatov2014,
 TITLE         = {Grain boundary energy function for fcc metals},
 AUTHOR        = {{Bulatov}, V. V. and {Reed}, B. W. and {Kumar}, M.},
 JOURNAL       = {Acta Materialia},
 VOLUME        = {65},
 PAGES         = {161},
 YEAR          = {2014}
}

@INBOOK{Herring1951,
   AUTHOR      = {{Herring}, C.},
   CHAPTER     = {8},
   PAGES       = {143},
   TITLE       = {The physics of powder metallurgy},
   PUBLISHER   = {McGraw-Hill},
   YEAR        = {1951}
}

@ARTICLE{Mason2017,
   AUTHOR      =  {{Mason}, J. K.},
   TITLE       =  {Stability and motion of arbitrary grain boundary junctions},
   JOURNAL     =  {Acta Materialia},
   VOLUME      =  {125},
   PAGES       =  {286},
   YEAR        =  {2017},
}

@ARTICLE{King1999,
   AUTHOR      =  {{King}, A. H.},
   TITLE       =  {The Geometric and Thermodynamic Properties of Grain Boundary Junctions},
   JOURNAL     =  {Interface Science},
   VOLUME      =  {7},
   PAGES       =  {251},
   YEAR        =  {1999},
}

@ARTICLE{Fortes1993,
   AUTHOR      =  {{Fortes}, M. A.},
   TITLE       =  {Stability and dissociation of quadruple junctions in polycrystals},
   JOURNAL     =  {Interface Science},
   VOLUME      =  {1},
   PAGES       =  {147},
   YEAR        =  {1993},
}

@ARTICLE{Taylor1976,
   AUTHOR      = {{Taylor}, J. E.},
   TITLE       = {The Structure of Singularities in Soap-Bubble-Like and Soap-Film-Like Minimal Surfaces},
   JOURNAL     = {Annals of Mathematics},
   VOLUME      = {103},
   NUMBER      = {3},
   PAGES       = {489},
   YEAR        = {1976},
}

@ARTICLE{Weaire1984,
   AUTHOR      = {{Weaire}, D. and {Rivier}, N.},
   TITLE       = {Soap, cells and statistics---random patterns in two dimensions},
   JOURNAL     = {Contemporary Physics},
   VOLUME      = {25},
   NUMBER      = {1},
   PAGES       = {59},
   YEAR        = {1984},
}

@ARTICLE{Glazier1992,
   AUTHOR      = {{Glazier}, J. A. and {Weaire}, D.},
   TITLE       = {The kinetics of cellular patterns},
   JOURNAL     = {Journal of Physics: Condensed Matter},
   VOLUME      = {4},   
   PAGES       = {1867},
   YEAR        = {1992},
}

@ARTICLE{Macpherson2007,
   AUTHOR      =  {{MacPherson}, R. D. and {Srolovitz}, D. J.},
   TITLE       =  {The von {N}eumann relation generalized to coarsening of three-dimensional microstructures},
   JOURNAL     =  {Nature},
   VOLUME      =  {446},
   PAGES       =  {1053},
   YEAR        =  {2007},
}

@ARTICLE{Matzke1945,
   AUTHOR      =  {{Matzke}, E. B.},
   TITLE       =  {The Three-Dimensional Shapes of Bubbles in Foams},
   JOURNAL     =  {PNAS},
   VOLUME      =  {31},
   NUMBER      =  {9},
   PAGES       =  {281},
   YEAR        =  {1945},
}

@ARTICLE{Rohrer2011,
   AUTHOR      =  {{Rohrer}, G. S.},
   TITLE       =  {Grain boundary energy anisotropy: a review},
   JOURNAL     =  {Journal of Materials Science},
   VOLUME      =  {46},
   PAGES       =  {5881},
   YEAR        =  {2011},
}

@ARTICLE{Marks2012,
   TITLE       =  {Equilibrium and stability of triple junctions in anisotropic systems},
   AUTHOR      =  {{Marks}, R. A. and {Glaeser}, A. M.},
   JOURNAL     =  {Acta Materialia},
   VOLUME      =  {60},
   PAGES       =  {349},
   YEAR        =  {2012},
}

@ARTICLE{Hallberg2019,
   TITLE       =  {Modeling of grain growth under fully anisotropic grain boundary energy},
   AUTHOR      =  {{Hallberg}, H. and {Bulatov}, V. V.},
   JOURNAL     =  {Modelling and Simulation in Materials Science and Engineering},
   VOLUME      =  {27},
   NUMBER      =  {4},
   PAGES       =  {045002},
   YEAR        =  {2019},
}

@ARTICLE{Hallberg2024b,
  AUTHOR       =  {{Hallberg}, H. and {Blixt}, K. H.},
  TITLE        =  {Assessing grain boundary variability through phase field crystal simulations},
  JOURNAL      =  {Physical Review Materials},
  VOLUME       =  {8},
  NUMBER       =  {3},
  PAGES        =  {113605},
  YEAR         =  {2024},
}

@ARTICLE{Gui2008,
   AUTHOR      =  {{Gui}, C. and {Schatzman}, M.},
   TITLE       =  {Symmetric Quadruple Phase Transitions},
   JOURNAL     =  {Indiana University Mathematics Journal},
   VOLUME      =  {57},
   NUMBER      =  {2},
   PAGES       =  {781},
   YEAR        =  {2008},
}

@ARTICLE{Reed2004,
   AUTHOR      =  {{Reed}, B. W. and {Minich}, R. W. and {Rudd}, R. E. and {Kumar}, M.},
   TITLE       =  {The structure of the cubic coincident site lattice rotation group},
   JOURNAL     =  {Acta Crystallographica A},
   VOLUME      =  {60},
   PAGES       =  {263},
   YEAR        =  {2004},
}

@ARTICLE{Stavans1993,
   AUTHOR      =  {{Stavans}, J.},
   TITLE       =  {The evolution of cellular structures},
   JOURNAL     =  {Reports on Progress in Physics},
   VOLUME      =  {56},
   PAGES       =  {733},
   YEAR        =  {1993},
}

@ARTICLE{Lazar2012,
   AUTHOR      =  {{Lazar}, E. A. and {Mason}, J. K. and {MacPherson}, R. D. and {Srolovitz}, D. J.},
   TITLE       =  {Complete Topology of Cells, Grains, and Bubbles in Three-Dimensional Microstructures},
   JOURNAL     =  {Physical Review Letters},
   VOLUME      =  {109},
   PAGES       =  {095505},
   YEAR        =  {2012},
}

@ARTICLE{Farhadifar2007,
   AUTHOR      =  {{Farhadifar}, R. and {R{\"o}per}, J. C. and {Aigouy}, B. and {Eaton}, S. and {J{\"u}licher}, F.},
   TITLE       =  {The influence of cell mechanics, cell-cell interactions, and proliferation on epithelial packing},
   JOURNAL     =  {Current Biology},
   VOLUME      =  {17},
   NUMBER      =  {24},
   PAGES       =  {2095},
   YEAR        =  {2007},
}

@ARTICLE{Hilgenfeldt2001,
   AUTHOR      =  {{Hilgenfeldt}, S. and {Kraynik}, A. M. and {Koehler}, S. A. and {Stone}, H. A.},
   TITLE       =  {An Accurate von Neumann’s Law for Three-Dimensional Foams},
   JOURNAL     =  {Physical Review Letters},
   VOLUME      =  {86},
   NUMBER      =  {12},
   PAGES       =  {2685},
   YEAR        =  {2001},
}

@article{herring1951some,
  title={Some theorems on the free energies of crystal surfaces},
  author={{Herring}, C.},
  journal={Physical Review},
  volume={82},
  number={1},
  pages={87},
  year={1951},
  publisher={APS}
}

@book{porter2021phase,
  title={Phase transformations in metals and alloys},
  author={{Porter}, D. A. and {Easterling}, K. E. and {Sherif}, M. Y.},
  year={2021},
  publisher={CRC press}
}

@incollection{Weaire2015,
    author = {{Weaire}, D. and {Hutzler}, S.},
    editor = {{Terentjev}, E. M. and {Weitz}, D. A.},
    title = {Foams},
    booktitle = {The Oxford Handbook of Soft Condensed Matter},
    publisher = {Oxford University Press},
    year = {2015},
}

@article{wolf1991structure,
  title={Structure and energy of general grain boundaries in bcc metals},
  author={{Wolf}, D},
  journal={Journal of Applied Physics},
  volume={69},
  number={1},
  pages={185},
  year={1991},
  publisher={American Institute of Physics}
}

@article{rohrer2007distribution,
  title={The distribution of grain boundary planes in polycrystals},
  author={{Rohrer}, G. S.},
  journal={JOM},
  volume={59},
  number={9},
  pages={38},
  year={2007},
  publisher={Springer}
}

@Article{Wirth2009,
  author  = {{Wirth}, K. and {Dreiding}, A. S.},
  title   = {Edge lengths determining tetrahedrons},
  journal = {Elemente der Mathematik},
  year    = {2009},
  volume  = {64},
  pages   = {160},
}

@ARTICLE{Hoffman1972,
   AUTHOR      = {{Hoffman}, D. W. and {Cahn}, J. W.},
   TITLE       = {A vector thermodynamics for anisotropic systems -- {I}. {F}undamentals and application to plane surface junctions},
   JOURNAL     = {Surface Science},
   VOLUME      = {31},
   PAGES       = {368},
   YEAR        = {1972},
}

@ARTICLE{CorderoBorboa2024,
   AUTHOR      = {{Cordero-Borboa}, A. E. and {Unda-Angeles}, R.},
   TITLE       = {Epifluorescence microscopy study of a quadruple node of triple junctions of grain boundaries in a Eu$^{2+}$-decorated highly textured composite of (Cl, Br)(K, Rb) and I(K, Rb) solid solutions},
   JOURNAL     = {Journal of Microscopy},
   VOLUME      = {296},
   PAGES       = {48},
   YEAR        = {2024},
}

@ARTICLE{Rhines1970,
   AUTHOR      =  {{Rhines}, F. N.},
   TITLE       =  {Geometry of Grain Boundaries},
   JOURNAL     =  {Metallurgical Transactions},
   VOLUME      =  {1},
   PAGES       =  {1105},
   YEAR        =  {1970},
}

@ARTICLE{Delaunay1934,
   AUTHOR      =  {{Delaunay}, B.},
   TITLE       =  {Sur la sph{\`e}re vide. {\`A} la m{\'e}moire de {G}eorges {V}orono{\"i}},
   JOURNAL     =  {Bulletin de l'Acad{\'e}mie des Sciences de l'URSS},
   NUMBER      =  {6},
   PAGES       =  {793},
   YEAR        =  {1934},
}
\bibliographystyle{sciencemag}


\paragraph*{Author contributions:}
Conceptualization: B.W.R. and V.V.B.; Methodology: H.H., V.V.B. and B.W.R.; Investigation: H.H., V.V.B. and B.W.R.; Visualization: H.H. and V.V.B.; Project administration: H.H.; Writing – original draft: H.H.; Writing – review and editing: H.H., V.V.B., B.W.R. and M.K.

\paragraph*{Competing interests:}
There are no competing interests to declare.

\paragraph*{Data, code and materials availability:}
All data needed to evaluate the conclusions are present in the paper and/or the Supplementary Materials. No physical materials were generated in this work.


\subsection*{Supplementary materials}
Materials and Methods\\
Supplementary Text\\
Figs. S1 to S7\\
Reference \textit{(\arabic{enumiv})}\\


\newpage


\renewcommand{\thefigure}{S\arabic{figure}}
\renewcommand{\thetable}{S\arabic{table}}
\renewcommand{\theequation}{S\arabic{equation}}
\renewcommand{\thepage}{S\arabic{page}}
\setcounter{figure}{0}
\setcounter{table}{0}
\setcounter{equation}{0}
\setcounter{page}{1}


\begin{center}
	\section*{Supplementary Materials for\\ \scititle}
	
	H{\aa}kan~Hallberg$^{1\ast}$,
	Vasily~V.~Bulatov$^{2}$,
	Bryan~W.~Reed$^{3}$,
	Mukul~Kumar$^{2}$\\
	\small$^{1}$Division of Solid Mechanics, Lund University, Lund, Sweden.\\
	\small$^{2}$Lawrence Livermore National Laboratory, Livermore, USA.\\
	\small$^{3}$Integrated Dynamic Electron Solutions, Inc., Pleasanton, USA.\\
	\small$^\ast$Corresponding author. Email: hakan.hallberg@solid.lth.se
\end{center}

\subsubsection*{This PDF file includes:}
Materials and Methods\\
Supplementary Text\\
Figures S1 to S7\\

\newpage


\subsection*{Materials and Methods}

\subsubsection*{4-node equilibrium and constructibility}\label{sect_4n}
Consider a 4-node at position $\bm{R}$, from which four 3-lines emanate, as illustrated in Fig.~\ref{fig_001}H in the main text. Let the four 3-lines intersect the boundary of a small tetrahedral control volume at the points defined by the vectors
\begin{equation}
    \bm{r}_{i} = l_{i}\hat{\bm{t}}_{i}, \quad i=1,2,3,4
	\label{eqnX1}
\end{equation}
originating at the 4-node position $\bm{R}$. In Eq.~(\ref{eqnX1}), $l_{i}=|\bm{r}_{i}|$ is the length of the 3-line segment $i$ inside the control volume and $\hat{\bm{t}}_{i}$ is the corresponding unit tangent, directed away from the 4-node. The six interface patches are indexed by pairs $ij$, $1\le i < j\le 4$ and the patch $ij$ is spanned locally by the two 3-line directions $\hat{\bm{t}}_{i}$ and $\hat{\bm{t}}_{j}$. The patch area is
\begin{equation}
	A_{ij} = \frac{1}{2}|\bm{r}_{i}\times\bm{r}_{j}| = \frac{1}{2}l_{i}l_{j}|\hat{\bm{t}}_{i}\times\hat{\bm{t}}_{j}|
	\label{eqnX2}
\end{equation}
and its unit normal is
\begin{equation}
	\hat{\bm{n}}_{ij} = \frac{\hat{\bm{t}}_{i}\times\hat{\bm{t}}_{j}}{|\hat{\bm{t}}_{i}\times\hat{\bm{t}}_{j}|}, \quad \hat{\bm{n}}_{ji} = -\hat{\bm{n}}_{ij}.
	\label{eqnX3}
\end{equation}
For each ordered pair $ij$, define also the in-plane unit vector
\begin{equation}
	\hat{\bm{m}}_{ij} = \hat{\bm{n}}_{ij}\times\hat{\bm{t}}_{i}.
	\label{eqnX4}
\end{equation}
The vector $\hat{\bm{m}}_{ij}$ lies in the interface plane $ij$, is perpendicular to 3-line $i$ and gives the direction of the capillary force exerted by interface $ij$ on 3-line $i$, according to the adopted sign convention. The vectors $\hat{\bm{n}}_{ij}$ and $\hat{\bm{m}}_{ij}$ are illustrated in Fig.~\ref{fig_001}I. The local excess interface energy in the control volume is then
\begin{equation}
    E = \sum_{i=1}^{3}\sum_{j=i+1}^{4} A_{ij}\gamma_{ij}(\hat{\bm{n}}_{ij}) .
	\label{eqn20}
\end{equation}
The subscript $ij$ includes all fixed interface-specific information, such as the identity of the adjacent cells or, for the boundaries of crystalline grains, the misorientation of the two grains. The local variation changes only the interface normal $\hat{\bm{n}}_{ij}$. A virtual perturbation gives
\begin{equation}
	\delta E = \sum_{i=1}^{3}\sum_{j=i+1}^{4}\left(\gamma_{ij}\delta A_{ij} + A_{ij}\delta\gamma_{ij}\right)
	\label{eqn21}
\end{equation}
with
\begin{equation}
	\delta\gamma_{ij} = \nabla_{\hat{\bm{n}}}\gamma_{ij}\cdot\delta\hat{\bm{n}}_{ij}.
	\label{eqn21b}
\end{equation}
Here $\nabla_{\hat{\bm{n}}}\gamma_{ij}$ denotes the derivative of the interface energy with respect to changes in the unit-normal direction $\hat{\bm{n}}_{ij}$. Since $\hat{\bm{n}}_{ij}$ is constrained to lie on the unit sphere, only the tangential, or surface-gradient, component of $\nabla_{\hat{\bm{n}}}\gamma_{ij}$ contributes to admissible inclination variations. The force per unit length exerted on 3-line $i$ by boundary $ij$ is
\begin{equation}
    \bm{\xi}_{ij} = \gamma_{ij}\hat{\bm{m}}_{ij} + \hat{\bm{n}}_{ij}\left(\hat{\bm{m}}_{ij}\cdot\nabla_{\hat{\bm{n}}}\gamma_{ij}\right).
	\label{eqnX5}
\end{equation}
Here, $\gamma_{ji}=\gamma_{ij}$, $\hat{\bm{n}}_{ji}=-\hat{\bm{n}}_{ij}$ and $\hat{\bm{m}}_{ji}=\hat{\bm{n}}_{ji}\times\hat{\bm{t}}_{j}$. The first term in Eq.~(\ref{eqnX5}) is the capillary contribution, and the second term is the torque contribution associated with the dependence of the interface energy on the boundary plane normal. The latter is the standard Cahn--Hoffman torque contribution
associated with the orientation dependence of interfacial free energy
\cite{Hoffman1972}. Thus, the Herring residual for 3-line $i$ is
\begin{equation}
	\bm{H}_{i} = \sum_{j\ne i}\bm{\xi}_{ij}, \quad i=1,2,3,4
	\label{eqnX6}
\end{equation}
The scale-invariant local mechanical equilibrium conditions for the 4-node are therefore
\begin{equation}
	\bm{H}_{i} = \bm{0}, \quad i=1,2,3,4
	\label{eqnX7}
\end{equation}
which, for 3-line, 1 yields
\begin{equation}
    \bm{H}_{1} = \bm{\xi}_{12} + \bm{\xi}_{13} + \bm{\xi}_{14} = \bm{0},
	\label{eqnX8}
\end{equation}
with analogous equations for 3-lines 2, 3 and 4. These are precisely the fully anisotropic Herring equations applied to the four 3-lines meeting at the 4-node.

The reason for writing the equilibrium conditions in the form Eq.~(\ref{eqnX7}) is scale invariance. The vectors $\bm{r}_{i}$ defining the boundary of the control volume are arbitrary, provided that the control volume is small enough that the interface patches can be treated as locally planar. Therefore, changing a length $l_{i}$ while keeping the local directions $\hat{\bm{t}}_{i}$ fixed cannot change the local equilibrium condition. Consequently, the force per unit length on each 3-line must vanish separately.

This also clarifies the relation between the four 3-line conditions and a direct variation of the 4-node position. If the 4-node point is displaced by $\delta\bm{R}$ while the control volume boundary points remain fixed, then
\begin{equation}
	\delta\bm{r}_{i} = -\delta\bm{R}, \quad i=1,2,3,4.
	\label{eqnX9}
\end{equation}
The corresponding first variation can be written as
\begin{equation}
	\delta E = -\left(\sum_{i=1}^{4}l_{i}\bm{H}_{i}\right)\cdot\delta\bm{R}
	\label{eqnX10}
\end{equation}
where $\bm{H}_{i}$ is defined as the force per unit length acting on the 3-line $i$, with the normalization adopted in Eq.~(\ref{eqnX6}). Thus, the force conjugate to a common displacement of the 4-node is a length-weighted linear combination of the four 3-line Herring residuals. Other equivalent normalizations of $\bm{H}_{i}$, or of the local patch areas, would only change a common numerical prefactor and would not affect the equilibrium condition. Hence, satisfaction of the four scale-invariant conditions $\bm{H}_{i}=\bm{0}$ implies vanishing of the direct force on the 4-node. Conversely, because the lengths $l_{i}$ are arbitrary cutout parameters, the vanishing of the weighted resultant for one particular choice of $l_{i}$ is not the fundamental local equilibrium condition. The local equilibrium condition is instead the simultaneous satisfaction of the four Herring equations in Eq.~(\ref{eqnX7}).

It follows from the preceding variation that there is no additional independent first-variation force-balance equation associated with the 4-node itself. The 4-node-specific problem is instead one of compatibility: whether the four Herring equations can be satisfied simultaneously by one common, non-degenerate local geometry. This also provides a natural definition of 4-node constructibility when the interface energies are fully anisotropic. With Eq.~(\ref{eqnX6}), a fully anisotropic 4-node is therefore constructible if there exists an admissible, non-degenerate set of four unit 3-line directions $\{\hat{\boldsymbol t}_1,\hat{\boldsymbol t}_2, \hat{\boldsymbol t}_3,\hat{\boldsymbol t}_4\}$ for which
\begin{equation}
    \bm{H}_{i} \left( \hat{\bm{t}}_{1}, \hat{\bm{t}}_{2}, \hat{\bm{t}}_{3}, \hat{\bm{t}}_{4} \right) = \bm{0}, \qquad i=1,2,3,4.
\end{equation}
If no such geometry exists, the 4-node is frustrated. Thus, constructibility under full anisotropy is an existence problem for the simultaneous solution of four coupled anisotropic Herring equations.

The semi-isotropic result follows as a special analytically solvable limit obtained when the interface energies are independent of interface inclination, $\nabla_{\hat{\boldsymbol n}}\gamma_{ij}=0$, and each $\gamma_{ij}$ is a fixed scalar for the interface $ij$. The four Herring residuals then reduce to
\begin{equation}
    \boldsymbol H_i = \sum_{j\ne i} \gamma_{ij}\hat{\boldsymbol m}_{ij} = \boldsymbol 0, \qquad i=1,2,3,4.
    \label{eqnX11}
\end{equation}
In this semi-isotropic limit, the six scalar energies can be identified with the six fixed edge lengths of a tetrahedral dual.

In the semi-isotropic setting, a necessary condition for tetrahedral constructibility can be obtained from the face areas of the tetrahedral dual. The area of the triangular face associated with 3-line $i$ can be evaluated by Heron's formula
\begin{equation}
	\mathcal{A}_{i} = \sqrt{s_{i}\left(s_{i}-\gamma_{ij}\right)\left(s_{i}-\gamma_{ik}\right)\left(s_{i}-\gamma_{il}\right)}, \quad s_{i} = \frac{1}{2}\left(\gamma_{ij}+\gamma_{ik}+\gamma_{il}\right), \quad j,k,l\ne i.
	\label{eqnX24}
\end{equation}
For a constructible tetrahedral dual, let $\hat{\bm{N}}_{i}$ denote the outward unit normal of face $i$. Since the integral of the unit normal over any closed surface vanishes, the area vectors of the four faces must satisfy
\begin{equation}
	\sum_{i=1}^{4}\mathcal{A}_{i}\hat{\bm{N}}_{i} = \bm{0}.
	\label{eqn45}
\end{equation}
It follows that each face area must satisfy
\begin{equation}
	\mathcal{A}_{i} \le \sum_{j\ne i}\mathcal{A}_{j}, \quad i=1,\ldots,4.
	\label{eqn46}
\end{equation}
Equality in Eq.~(\ref{eqn46}) corresponds to a degenerate limiting case in which the area vectors are collinear. A non-degenerate tetrahedron therefore requires strict satisfaction of the area inequalities, in addition to the Cayley--Menger determinant being positive. The inequalities in Eq.~(\ref{eqn46}) are the area-vector analogue of the triangle inequality for a 3-node. They are necessary for foldability of the tetrahedral dual, but they are not sufficient: the six edge lengths contain compatibility information that is not captured by the four face areas alone. Therefore, $\Delta_{\text{CM}}$ remains the appropriate constructibility criterion for the tetrahedral dual.

\subsubsection*{4-node constructibility based on corner-angle foldability}
The main text classifies frustrated cutouts by counting triangle-type corner failures after non-constructibility has already been established by $\Delta_{\rm CM}<0$. Here we give the complete corner-foldability condition, including the angle-sum condition, to justify that this classification is complete and to connect the F3 class of frustrated 4-nodes to violation of an opposite face-area inequality. This formulation is closely related to the classical edge-length problem for tetrahedra, in which a facial sextuple defines four triangular faces but is tetrahedral only if the corresponding net can be folded into a non-degenerate tetrahedron \cite{Wirth2009}.

\begin{figure}
	\centering
	\includegraphics[width=0.25\textwidth]{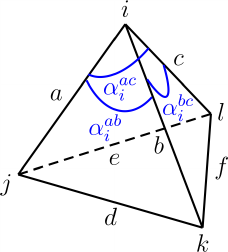}
	\caption{\textbf{Quantities used to establish the corner-angle foldability criterion for a 4-node.} The vertices of the tetrahedral dual are indexed by $i,j,k,l$ and the tetrahedral edge lengths, corresponding to the energies of the interfaces meeting at the 4-node, are denoted by $a,b,c,d,e,f$. The corner angles at vertex $i$ are shown in the figure as $\alpha_{i}^{ab},\alpha_{i}^{ac},\alpha_{i}^{bc}$. Corner angles at the other vertices are defined in the same way.}
	\label{fig16}
\end{figure}

Using the notation for the tetrahedral dual illustrated in Fig.~\ref{fig16}, the tetrahedron vertices are indexed by $i,j,k,l$, and the edge lengths are denoted by
\begin{equation}
	\begin{array}{l}
		\vspace{2mm}
		\displaystyle a=\gamma_{ij}, \quad b=\gamma_{ik}, \quad c=\gamma_{il}, \\
		\displaystyle d=\gamma_{jk}, \quad e=\gamma_{jl}, \quad f=\gamma_{kl}.
	\end{array}
	\label{eqnX60}
\end{equation}
The three corner angles at vertex $i$ are determined by the three adjacent triangular faces:
\begin{equation}
	\begin{array}{l}
		\vspace{2mm}
		\displaystyle \alpha_{i}^{ab} =  \cos^{-1}\left(\frac{a^{2}+b^{2}-d^{2}}{2ab}\right), \\
		\vspace{2mm}
		\displaystyle \alpha_{i}^{ac} = \cos^{-1}\left(\frac{a^{2}+c^{2}-e^{2}}{2ac}\right), \\
		\displaystyle \alpha_{i}^{bc} =
		\cos^{-1}\left(\frac{b^{2}+c^{2}-f^{2}}{2bc}\right).
	\end{array}
	\label{eqnX61}
\end{equation}
These angles satisfy $0<\alpha_{i}^{ab},\alpha_{i}^{ac},\alpha_{i}^{bc}<\pi$. The same construction gives three corner angles at each of the other vertices. A first 4-node constructibility requirement is that all four triangular faces satisfy the ordinary triangle inequalities for each face, e.g., $\gamma_{ij}\le\gamma_{ik}+\gamma_{jk}$ and cyclic permutations, so that these angles are real. For the angles in Eq.~(\ref{eqnX61}) to form a non-degenerate trihedral corner, they must satisfy
\begin{subequations}\label{eqnX46}
	\begin{align}
		\displaystyle \alpha_{i}^{ab} < \alpha_{i}^{ac} + \alpha_{i}^{bc}, \label{eqnX46-a} \\
		\displaystyle \alpha_{i}^{ac} < \alpha_{i}^{ab} + \alpha_{i}^{bc}, \label{eqnX46-b} \\
		\displaystyle \alpha_{i}^{bc} < \alpha_{i}^{ab} + \alpha_{i}^{ac}, \label{eqnX46-c} \\
		\displaystyle \alpha_{i}^{ab} + \alpha_{i}^{ac} + \alpha_{i}^{bc} < 2\pi \label{eqnX46-d}.
	\end{align}
\end{subequations}
The first three conditions are the triangle-type inequalities. These are the inequalities used in the main text to classify frustrated cutouts into the classes F3 and F4. The fourth condition is the angle-sum condition; it excludes a flat or overfolded corner. Although this condition is not needed as a separate classifier in the main text, it is needed here to prove that the number of triangle-type failures can only be three or four and to identify the complementary failure mode in F3 cases. A slightly less visual variant of the conditions in Eq.~(\ref{eqnX46}) is obtained if one constructs the Gram matrix
\begin{equation}
	\bm{G}_{i} = \left[\begin{array}{ccc}
		1 & \cos{\alpha_{i}^{ab}} & \cos{\alpha_{i}^{ac}} \\
		\cos{\alpha_{i}^{ab}} & 1 & \cos{\alpha_{i}^{bc}} \\
		\cos{\alpha_{i}^{ac}} & \cos{\alpha_{i}^{bc}} & 1
	\end{array}\right].
	\label{eqnX47}
\end{equation}
The matrix $\bm{G}_{i}$ relates to vertex $i$, but is constructed identically for all four vertices. Non-degenerate foldability requires $\det\bm{G}_{i}>0$ or, explicitly, 
\begin{equation}
	\det\bm{G}_{i} = 1 + 2\cos{\alpha_{i}^{ab}}\cos{\alpha_{i}^{ac}}\cos{\alpha_{i}^{bc}} - \cos^{2}\alpha_{i}^{ab} - \cos^{2}\alpha_{i}^{ac} - \cos^{2}\alpha_{i}^{bc} > 0.
	\label{eqnX48}
\end{equation}
Since the squared volume of the tetrahedron can be written in terms of the scalar triple product of the three edge vectors meeting at vertex $i$, one obtains
\begin{equation}
	\Delta_{\text{CM}} = 8a^{2}b^{2}c^{2}\det\bm{G}_{i},
	\label{eqnX51}
\end{equation}
with $\Delta_{\text{CM}}$ being defined in Eq.~(\ref{eqn_005}). Thus, provided the four triangular faces are individually constructible, the Cayley--Menger criterion is equivalent to the requirement that the three face angles meeting at any vertex define a valid trihedral angle. Since the prefactor in Eq.~(\ref{eqnX51}) is positive, the sign of $\det\bm{G}_{i}$ is the same as the sign of $\Delta_{\text{CM}}$ for every choice of vertex $i$. However, the particular inequality in Eq.~(\ref{eqnX46}) that fails may differ among vertices. The pattern of triangle-type and angle-sum failures provides a geometric classification of non-foldable cutouts.

\subsubsection*{Restriction on the number of triangle-type corner failures}
Consider an edge-matched tetrahedral cutout whose four triangular faces are strictly constructible. The three corner angles at vertex $i$ are given by Eq.~(\ref{eqnX61}) and the complete trihedral corner condition at vertex $i$ is provided by Eq.~(\ref{eqnX46}). Assume now that the four triangular faces are strictly constructible but that the tetrahedral dual is non-constructible. Then $\Delta_{\text{CM}}$ is negative. Since Eq.~(\ref{eqnX51}) holds at every vertex $i$, it follows that $\det(\bm{G}_{i})<0$, $i=1,2,3,4$. Thus, the complete trihedral corner condition fails at every vertex.

At a given vertex, the possible local failure modes are restricted. Since $\alpha_{i}^{ab},\alpha_{i}^{ac},\alpha_{i}^{bc}\in(0,\pi)$, at most one of the three triangle-type
inequalities can fail. Indeed, if for example
\begin{equation}
	\alpha_{i}^{ab}\geq \alpha_{i}^{ac}+\alpha_{i}^{bc} \quad \text{and} \quad
	\alpha_{i}^{ac}\geq \alpha_{i}^{ab}+\alpha_{i}^{bc}
\end{equation}
then addition gives
\begin{equation}
	\alpha_{i}^{ab}+\alpha_{i}^{ac}\geq\alpha_{i}^{ab}+\alpha_{i}^{ac}+2\alpha_{i}^{bc} ,
\end{equation}%
which implies $\alpha_{i}^{bc}\leq 0$, contradicting $\alpha_{i}^{bc}>0$. The same argument applies to any pair of triangle-type inequalities.

Moreover, a triangle-type failure and an angle-sum failure cannot occur simultaneously at the same vertex. If, for example, $\alpha_{i}^{ab}\geq \alpha_{i}^{ac}+\alpha_{i}^{bc}$ then
\begin{equation}
	\alpha_{i}^{ab}+\alpha_{i}^{ac}+\alpha_{i}^{bc}\leq2\alpha_{i}^{ab} .
\end{equation}
Since $\alpha_{i}^{ab}<\pi$, this gives
\begin{equation}
	\alpha_{i}^{ab}+\alpha_{i}^{ac}+\alpha_{i}^{bc}<2\pi .
\end{equation}
Thus, the angle-sum condition is satisfied whenever one of the triangle-type inequalities fails. Conversely, if the angle-sum condition fails, $\alpha_{i}^{ab}+\alpha_{i}^{ac}+\alpha_{i}^{bc}\geq2\pi$, then, for example,
\begin{equation}
	\alpha_{i}^{bc}+\alpha_{i}^{ac}=\alpha_{i}^{ab}+\alpha_{i}^{ac}+\alpha_{i}^{bc}-\alpha_{i}^{ab}\geq2\pi-\alpha_{i}^{ab} .
\end{equation}
Since $\alpha_{i}^{ab}<\pi$, we have $2\pi-\alpha_{i}^{ab}>\alpha_{i}^{ab}$, and hence $\alpha_{i}^{ac}+\alpha_{i}^{bc}>\alpha_{i}^{ab}$. The same cyclic argument shows that all three triangle-type inequalities are satisfied. Therefore, at each vertex, failure of the complete trihedral condition occurs in exactly one of two ways: either one triangle-type inequality fails or all three triangle-type inequalities hold while the angle-sum condition fails.

It remains to show that angle-sum failure can occur at no more than one vertex. Let
\begin{equation}
	S_{i}=\alpha_{i}^{ab}+\alpha_{i}^{ac}+\alpha_{i}^{bc}
\end{equation}
be the sum of the three corner angles at vertex $i$. Summing $S_{i}$ over the four vertices counts the angles of each triangular face exactly once. Since each of the four faces is a planar triangle, its three angles sum up to $\pi$. Hence
\begin{equation}
	\sum_{i=1}^{4}S_{i}=4\pi .
\end{equation}
If two vertices, say $p$ and $q$, both failed by the angle-sum condition, then
\begin{equation}
	S_{p}\geq 2\pi \quad \text{and} \quad S_{q}\geq 2\pi.
\end{equation}
This would imply $S_{p}+S_{q}\geq4\pi$. Since the remaining two angle sums are strictly positive, it follows that $\sum_{i=1}^{4}S_{i}>4\pi$, which contradicts the identity above. Therefore, at most one vertex can fail through the angle-sum condition.

Since the complete trihedral corner condition fails at all four vertices, and since at most one vertex can fail through the angle-sum condition, at least three vertices must fail through one of the triangle-type inequalities. Thus, for a non-constructible edge-matched tetrahedral cutout with all four triangular faces strictly constructible, the triangle-type failure count can only be three or four. These are the F3 and F4 classes used in the main text. In F3, the remaining vertex has no triangle-type failure; its failure is instead the angle-sum failure described by Eq.~(\ref{eqnX46-d}).

\subsubsection*{Connection between angle-sum failure and the opposite face-area inequality}
We next show that the unique F3 vertex without a triangle-type failure corresponds to violation of the opposite face-area inequality. Consider the tetrahedral cutout with vertices $i,j,k,l$, shown in Fig.~\ref{fig16}. The three corner angles at vertex $i$ are denoted by
\begin{equation}
	\alpha_{i}^{ab} = \angle jik,\qquad
	\alpha_{i}^{ac} = \angle jil,\qquad
	\alpha_{i}^{bc} = \angle kil .
\end{equation}
Thus,
\begin{equation}
	\begin{array}{l}
		\vspace{2mm}
		\displaystyle d^2 = a^2+b^2 - 2ab\cos\alpha_{i}^{ab}, \\
		\vspace{2mm}
		\displaystyle e^2 = a^2+c^2 - 2ac\cos\alpha_{i}^{ac}, \\
		\displaystyle f^2 = b^2+c^2 - 2bc\cos\alpha_{i}^{bc} .
	\end{array}
\end{equation}
Let
\begin{equation}
	S=\alpha_{i}^{ab}+\alpha_{i}^{ac}+\alpha_{i}^{bc} .
\end{equation}
The areas of the three faces incident to vertex $i$ are
\begin{equation}
	A_{ijk}=\frac{1}{2} ab\sin\alpha_{i}^{ab}, \quad
	A_{ijl}=\frac{1}{2} ac\sin\alpha_{i}^{ac}, \quad
	A_{ikl}=\frac{1}{2} bc\sin\alpha_{i}^{bc} .
\end{equation}
Let $A_{jkl}$ denote the area of the face opposite vertex $i$, whose side lengths are $d,e,f$. Define
\begin{equation}
	T_{i} = A_{ijk} + A_{ijl} + A_{ikl}.
\end{equation}
We now compare $A_{jkl}$ with $T_i$. By Heron's formula or, equivalently, by the standard expression for the squared area of a triangle in terms of its side lengths,
\begin{equation}
	16A_{jkl}^2 = 2d^2e^2 + 2e^2f^2 + 2f^2d^2 - d^4 - e^4 - f^4 .
\end{equation}
Substituting the law-of-cosines expressions for $d^2,e^2,f^2$, and subtracting $T_i^2$, gives the identity
\begin{equation}
	\begin{array}{lcl}
		\vspace{2mm}
		\displaystyle A_{jkl}^2-T_i^2 &=& \displaystyle -abc\sin\frac{S}{2}
		\left[a\sin\frac{\alpha_{i}^{ab}+\alpha_{i}^{ac}-\alpha_{i}^{bc}}{2} + \right.\\
		&& \displaystyle \left. b\sin\frac{\alpha_{i}^{ab}+\alpha_{i}^{bc}-\alpha_{i}^{ac}}{2} +
		c\sin\frac{\alpha_{i}^{ac}+\alpha_{i}^{bc}-\alpha_{i}^{ab}}{2}\right].
	\end{array}
	\label{eqnX71}
\end{equation}
This identity is purely algebraic and follows only from the six edge lengths of the edge-matched cutout.

Assume now that the three triangle-type corner inequalities at vertex $i$ are satisfied, cf. Eq.~(\ref{eqnX46-a})--(\ref{eqnX46-c}). Then
\begin{equation}
	\frac{\alpha_{i}^{ab}+\alpha_{i}^{ac}-\alpha_{i}^{bc}}{2}>0,\qquad
	\frac{\alpha_{i}^{ab}+\alpha_{i}^{bc}-\alpha_{i}^{ac}}{2}>0,\qquad
	\frac{\alpha_{i}^{ac}+\alpha_{i}^{bc}-\alpha_{i}^{ab}}{2}>0 .
\end{equation}
Since the face triangles are strictly constructible, the corner angles satisfy $0<\alpha_{i}^{ab},\alpha_{i}^{ac},\alpha_{i}^{bc}<\pi$. Consequently, all three sine factors in the square brackets in Eq.~(\ref{eqnX71}) are positive, and the whole bracket is positive.

If vertex $i$ fails by the angle-sum condition, then
\begin{equation}
	S = \alpha_{i}^{ab}+\alpha_{i}^{ac}+\alpha_{i}^{bc}>2\pi .
\end{equation}
Since each of $\alpha_{i}^{ab},\alpha_{i}^{ac},\alpha_{i}^{bc}$ is less than $\pi$, we also have $S<3\pi$. Therefore
\begin{equation}
	\sin\frac{S}{2}<0 .
\end{equation}
It follows from the identity above that
\begin{equation}
	A_{jkl}^2-T_i^2>0 .
\end{equation}
Since both $A_{jkl}$ and $T_{i}$ are positive, this is equivalent to $A_{jkl}>T_{i}$. That is,
\begin{equation}
	A_{jkl} > A_{ijk} + A_{ijl} + A_{ikl}.
	\label{eqnX70}
\end{equation}
Thus, if the unique failure at vertex $i$ is the angle-sum failure, then the face opposite vertex $i$ violates the scalar face-area inequality. Conversely, under the same assumption that the three triangle-type corner inequalities hold, the bracket in the identity is positive. Hence Eq.~(\ref{eqnX70}) implies
\begin{equation}
	A_{jkl}^2-T_i^2>0,
\end{equation}
which requires
\begin{equation}
	\sin\frac{S}{2}<0 .
\end{equation}%
Since $0<S<3\pi$, this gives $S>2\pi$. Therefore, when the three triangle-type corner inequalities hold, violation of the opposite face-area inequality is equivalent to failure of the angle-sum condition at that vertex.

Consequently, in a non-constructible edge-matched tetrahedral cutout with all four triangular faces strictly constructible, there are two possible cases: either one triangle-type corner inequality fails at all four vertices, or the triangle-type inequalities fail at three vertices while the remaining vertex fails only through the angle-sum condition. In the latter case, the face opposite the angle-sum-failure vertex satisfies Eq.~(\ref{eqnX70}), so the corresponding face-area margin is negative.


\subsection*{Supplementary Text}


\subsubsection*{Non-constructible regions in the tetrahedral-dual energy space}\label{sect_unstable_regions}
To further quantify the size of the region of non-constructibility identified in the previous section, sampling is performed in the six-dimensional tetrahedral-dual energy space associated with a semi-isotropic 4-node. The six interface energies are sampled randomly and independently from a bounded interval defined by
\begin{equation}
	\gamma_{ij} \in [1-\delta, 1+\delta], \quad 0 \le \delta <1
	\label{eqn55}
\end{equation}
where $\delta$ is the half-width of the sampled energy interval and therefore controls the admissible energy variation. Since both the triangle inequalities and the tetrahedral constructibility condition are homogeneous in the interface energies, the absolute energy scale is immaterial. The interval in eq.~(\ref{eqn55}) therefore defines a dimensionless measure of relative energy variation about the isotropic state $\gamma_{ij}=1$. For each value of $\delta$, the energy domain is the six-dimensional space $(\gamma_{12},\gamma_{13},\gamma_{14},\gamma_{23},\gamma_{24},\gamma_{34})$ and sampling is performed uniformly in this domain. This means that, for a particular value of $\delta$, each of the $N_{\text{s}}$ samples consists of six values of $\gamma_{ij}$ drawn independently from this domain. The $\delta$-dependent sampling described here uses $N_{\text{s}}=10^6$ samples at each value of $\delta$. The classification statistics reported at $\delta=0.8$ in Fig.~\ref{fig_003}A and Fig.~\ref{fig_003}C are obtained from a separate $N_{\text{s}}=2\times10^6$ sample. The constructibility of the 3-line force triangles defined by these interface energy values is evaluated by using the ordinary triangle inequalities for the four tetrahedral-dual faces and the constructibility of the corresponding 4-node is evaluated by $\Delta_{\text{CM}}$ in eq.~(\ref{eqn_005}). 

\begin{figure}
	\centering
	\includegraphics[width=0.9\textwidth]{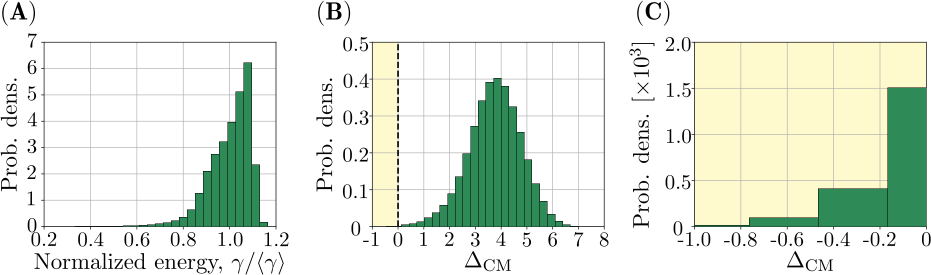}
	\caption{\textbf{GB5DOF-based energy sampling of the tetrahedral-dual constructibility criterion.} (\textbf{A}) Distribution of normalized scalar interface energies, $\gamma/\langle\gamma\rangle$, obtained by sampling random grain orientations in $SO(3)$ and random interface plane normals on the unit sphere, followed by evaluation of the corresponding GB5DOF energy. (\textbf{B}) Distribution of the Cayley--Menger determinant, $\Delta_{\text{CM}}$, for six-energy sets drawn independently from the normalized GB5DOF energy distribution and tested using the tetrahedral-dual constructibility criterion. The dashed vertical line indicates the 4-node constructibility limit. (\textbf{C}) A magnification of the small negative-determinant tail, indicated by the shaded region in (\textbf{B}).}
	\label{fig10}
\end{figure}

\begin{figure}
	\centering
	\includegraphics[width=\textwidth]{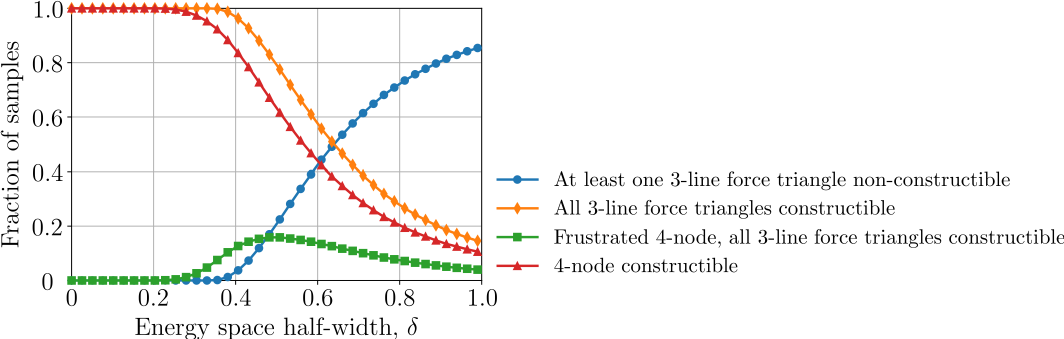}
	\caption{\textbf{Relative occurrence of 3-line force-triangle and 4-node constructibility.} The fractions of the sampled energy space, cf. eq.~(\ref{eqn55}), corresponding to 3-line force-triangle non-constructibility, 4-node constructibility and 4-node non-constructibility despite constructibility of all four constituent 3-line force triangles.}
	\label{fig8}
\end{figure}

Fig.~\ref{fig8} shows, as functions of $\delta$, the fractions of samples corresponding to the different constructibility scenarios. The corresponding conditional probability $P(\delta)$ is shown in Fig.~\ref{fig_003}A of the main text. The probability is evaluated as the number of samples in which the 4-node is non-constructible and the constituent 3-line force triangles are constructible, divided by the number of samples in which all four 3-line force triangles are constructible.

The results show that for small energy variation, approximately $\delta\le 0.25$, nearly all sampled configurations have constructible 3-line force triangles and 4-nodes. With increasing $\delta$, configurations with at least one non-constructible 3-line force triangle become increasingly common. Importantly, there is also a finite intermediate range in which all four 3-line force triangles are constructible while the 4-node is non-constructible. This demonstrates that the non-constructibility identified in the one-parameter example in the main text is not an isolated pathological case, but occupies a finite measure in the tetrahedral-dual energy space. The conditional probability in Fig.~\ref{fig_003}A increases with $\delta$, indicating that among configurations for which all four triangular faces are constructible, a non-negligible fraction nevertheless fails to form a constructible tetrahedral dual.

\subsubsection*{Opposite-edge structure of F4 failures}
Each triangle-type failure at a vertex can be associated with the tetrahedral edge opposite the failed corner angle. Using the edge notation introduced in Eq.~(\ref{eqnX60}) and Fig.~\ref{fig16}, the three F4 mode patterns observed in the random sampling map to the following opposite-edge pairs:
\begin{equation}
	\begin{array}{lcl}
		\vspace{2mm}
		\displaystyle \mathrm{tri1|tri2|tri2|tri3} & \mapsto & \gamma_{14}|\gamma_{23}, \\
		\vspace{2mm}
		\displaystyle \mathrm{tri2|tri1|tri3|tri2} & \mapsto & \gamma_{13}|\gamma_{24}, \\
		\displaystyle \mathrm{tri3|tri3|tri1|tri1} & \mapsto & \gamma_{12}|\gamma_{34}.
	\end{array}
\end{equation}
The four entries in each pattern correspond to vertices 1,2,3,4, and tri$k$ denotes failure of the $k$-th triangle-type inequality in the local corner-angle ordering. Thus, each F4 cutout identifies two opposite tetrahedral edges, each appearing twice among the four failed corner inequalities. In the $N_{\text{s}}=2\times10^6$ sample at $\delta=0.8$, all 134,855 F4 cases had this structure (Fig.~\ref{fig_003}C). The offending pair was exactly the two largest energies in 39.0\% of F4 cases, both offending edges were among the three largest in 78.7\%, and both were among the four largest in 97.5\%. The mean length of the offending pair was 1.48 times the mean length of the other four edges.

\subsubsection*{GB5DOF-based sampling of the tetrahedral-dual energy space}
The sampling in the previous section was intentionally independent of any particular interface energy model. While this provides a direct measure of regions of non-constructibility in the tetrahedral-dual energy space, it does not indicate whether such regions are populated by a realistic interface energy distribution. As a qualitative comparison, a second sampling is therefore performed using scalar interface energies obtained from the GB5DOF energy function \cite{Bulatov2014}. This provides a numerically generated distribution of interface energy values associated with random misorientations and boundary-plane inclinations, while retaining the same tetrahedral-dual constructibility tests used above. GB5DOF has, for example, been used together with a level set formulation to trace the evolution of grain boundaries under fully anisotropic interface energies \cite{Hallberg2019}.

For each sampled interface, two crystal orientations are drawn uniformly from $SO(3)$ and an interface plane normal is drawn uniformly on the unit sphere. The resulting interface energies, obtained from GB5DOF, are normalized by their mean value,
\begin{equation}
	\tilde{\gamma}=\frac{\gamma}{\langle\gamma\rangle}
	\label{eqn58}
\end{equation}
so that only relative energy variations enter the constructibility analysis.

A set of six normalized interface energies is drawn independently from this sampled distribution and assigned to the six edges of the tetrahedral dual, $(\tilde{\gamma}_{12},\tilde{\gamma}_{13},\tilde{\gamma}_{14},\tilde{\gamma}_{23},\tilde{\gamma}_{24},\tilde{\gamma}_{34})$. For each such six-energy set, the four constituent 3-line force triangles are tested using the triangle inequality, and 4-node constructibility is assessed using $\Delta_{\text{CM}}$.

It should be emphasized that this procedure, as well as the independent sampling above, does not represent a fully compatible four-grain 4-node geometry. The six interface energies are sampled independently and are not generated from one common set of four grain orientations and six geometrically compatible interface inclinations.

The sampled interface energy distribution is shown in Fig.~\ref{fig10}A. The distribution is concentrated near the mean energy, but contains a finite low-energy tail that generates occasional large contrasts among the six edge lengths of the tetrahedral dual. The resulting $\Delta_{\text{CM}}$ distribution is shown in Fig.~\ref{fig10}B. Most sampled six-energy sets produce positive determinants and therefore correspond to constructible tetrahedra. However, Fig.~\ref{fig10}C provides a zoom of the shaded region in Fig.~\ref{fig10}B (where $\Delta_{\text{CM}}$ is negative), showing a small negative tail. This indicates that a small fraction of the sampled six-energy sets satisfies the four 3-line force-triangle inequalities while failing the 4-node constructibility criterion. Thus, the 4-node non-constructibility region is sparsely, but not negligibly, populated by this GB5DOF-based energy distribution.

In the present sampling, $N_{\text{s}}=10^{6}$ six-energy sets were drawn from the normalized GB5DOF energy distribution. The fraction of all six-energy sets for which all 3-line force triangles are constructible while $\Delta_{\text{CM}}$ is negative is $3.2\times10^{-4}$. Since nearly all samples, 99.9\%, met the four 3-line force-triangle constructibility conditions, the corresponding conditional probability $P$, evaluated as in the previous section, is nearly identical, $P=3.2\times10^{-4}$.

As emphasized above, this result should not be interpreted as a prediction of the frequency of non-constructible 4-nodes in a real polycrystal. The six energies are sampled independently from the scalar GB5DOF energy distribution and are not generated from a compatible four-grain configuration.

\subsubsection*{Distance to the constructible boundary}
To evaluate the distance $d_{\log}$ defined in Eq.~(\ref{eqn_009}) of the main text, we numerically determine, for each frustrated energy sextuplet $\bm{\gamma}$, the closest trial sextuplet $\bm{\gamma}'$ on the constructible boundary. The minimization is performed in logarithmic energy coordinates and is restricted to trial sextuplets for which all four face-triangle inequalities are satisfied:
\begin{equation}
	d_{\log}(\bm{\gamma})	=
	\min_{\bm{\gamma}'}
	\left[
	\sum_{i<j}
	\left(
	\log\frac{\gamma'_{ij}}{\gamma_{ij}}
	\right)^2
	\right]^{1/2},
	\label{eqn_X42}
\end{equation}
subject to
\begin{equation}
	\Delta_{\text{CM}}(\bm{\gamma}')=0
	\label{eqn_X43}
\end{equation}
and to satisfaction of all four face-triangle inequalities. Thus, the optimization is restricted to the portion of the $\Delta_{\text{CM}}=0$ hypersurface that bounds the physically admissible region of tetrahedral-dual energy space.

For the representative frustrated samples shown in Fig.~\ref{fig_003}B, $d_{\log}$ correlates strongly with the normalized Cayley--Menger depth $\chi$. For comparison, $\chi$ also correlates strongly with the minimum residual $\Phi_{\min}$ defined below, although evaluation of $\Phi_{\min}$ requires optimization over 3-line directions together with a non-degeneracy constraint.

\subsubsection*{Supplementary minimum-residual construction of non-constructible 4-nodes}\label{sect_min_Herring}
\begin{figure}
	\centering
	\includegraphics[width=0.6\textwidth]{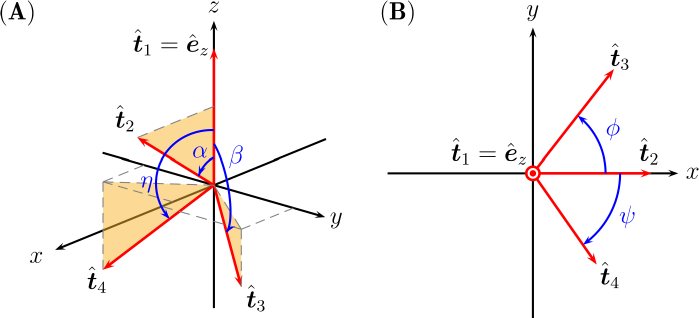}
	\caption{\textbf{Five-parameter representation of the four 3-line directions used in the minimum-residual construction.} A common rigid rotation is removed by setting $\hat{\bm{t}}_{1}=(0,0,1)$ and placing $\hat{\bm{t}}_{2}$ in the $xz$-plane. The remaining directions are parameterized according to Eq.~(\ref{eqnX20}). (\textbf{A}) Polar angles $\alpha$, $\beta$ and $\eta$, measured from $\hat{\bm{t}}_{1}$. (\textbf{B}) Azimuthal angles $\phi$ and $\psi$, measured in the $xy$-plane from the positive $x$-axis. The azimuth of $\hat{\bm{t}}_{2}$ is zero by construction.}
	\label{fig13}
\end{figure}

The distance $d_{\log}$ used in the main text measures how far a frustrated energy sextuplet lies from the constructible boundary. A complementary, more mechanical but also more algorithmic measure can be obtained by minimizing the four Herring residuals directly over the 3-line directions. This construction is useful for illustrating how incompatibility is distributed among the four 3-lines, but it requires a non-degeneracy constraint and is therefore kept as a supplementary characterization. Such a construction can be formulated without introducing arbitrary cutout lengths $l_{i}$. Instead of displacing the 4-node for a prescribed control volume, as in the previous section, the four scale-invariant Herring residuals $\bm{H}_{i}$ can be minimized directly over the possible 3-line directions. In the semi-isotropic case, Eqs.~(\ref{eqnX3}), (\ref{eqnX4}) and (\ref{eqnX11}) give $\bm{H}_{i}\left(\hat{\bm{t}}_{1},\hat{\bm{t}}_{2},\hat{\bm{t}}_{3},\hat{\bm{t}}_{4}\right)$ from which a dimensionless residual measure can be defined as
\begin{equation}
	\Phi\left(\hat{\bm{t}}_{1},\hat{\bm{t}}_{2},\hat{\bm{t}}_{3},\hat{\bm{t}}_{4}\right) = \frac{\sum_{i=1}^{4}|\bm{H}_{i}|^{2}}{\sum_{i=1}^{4}\left(\sum_{j\ne i}\gamma_{ij}\right)^{2}}.
	\label{eqnX18}	
\end{equation}
The denominator only fixes the scale to make $\Phi$ dimensionless and invariant under a common rescaling of all interface energies. The minimum-residual construction is obtained from
\begin{equation}
	\Phi_{\text{min}} = \min_{\hat{\bm{t}}_{1,2,3,4}}\Phi\left(\hat{\bm{t}}_{1},\hat{\bm{t}}_{2},\hat{\bm{t}}_{3},\hat{\bm{t}}_{4}\right).
	\label{eqnX19}	
\end{equation}
Four unit vectors contain eight angular degrees of freedom, while a common rotation accounts for three, leaving five independent degrees of freedom. These are represented by setting
\begin{equation}
	\begin{array}{ll}
		\vspace{2mm}
		\displaystyle \hat{\bm{t}}_{1} = \left(0,0,1\right), &
		\hat{\bm{t}}_{2} = \left(\sin\alpha,0,\cos\alpha\right), \\
		\displaystyle\hat{\bm{t}}_{3} = \left(\sin\beta\cos\phi,\sin\beta\sin\phi,\cos\beta\right), &
		\hat{\bm{t}}_{4} = \left(\sin\eta\cos\psi,\sin\eta\sin\psi,\cos\eta\right)
	\end{array}
	\label{eqnX20}
\end{equation}
and are further illustrated in Fig.~\ref{fig13}. For a constructible 4-node, $\Phi_{\text{min}}=0$, up to numerical tolerance. For a non-constructible 4-node, no non-degenerate set of directions can make all four residuals vanish simultaneously. In the numerical examples below, near-degenerate configurations are monitored using
\begin{equation}
	\rho_{ij} = |\hat{\bm{t}}_{i}\times\hat{\bm{t}}_{j}|.
	\label{eqnX23}
\end{equation}
A lower bound $\rho_{ij}\ge\rho_{0}$, with $\rho_{0}=0.1$, is imposed as a non-degeneracy constraint. Small values of $\rho_{ij}$ therefore indicate that the minimum-residual construction is approaching a collapse-like limit. The parameter $\rho_{0}$ is used only to regularize the numerical construction and to prevent complete collapse of an interface patch. It is not a physical parameter. Geometrically, if the optimized directions are embedded in a fixed local cutout, Eq.~(\ref{eqnX2}) gives $A_{ij}=\frac{1}{2}l_{i}l_{j}\rho_{ij}$. Thus, $\rho_{ij}\rightarrow 0$ corresponds to collapse of interface patch $ij$ and collinearity of its two bounding 3-lines.

\begin{figure}
	\centering
	\includegraphics[width=0.7\textwidth]{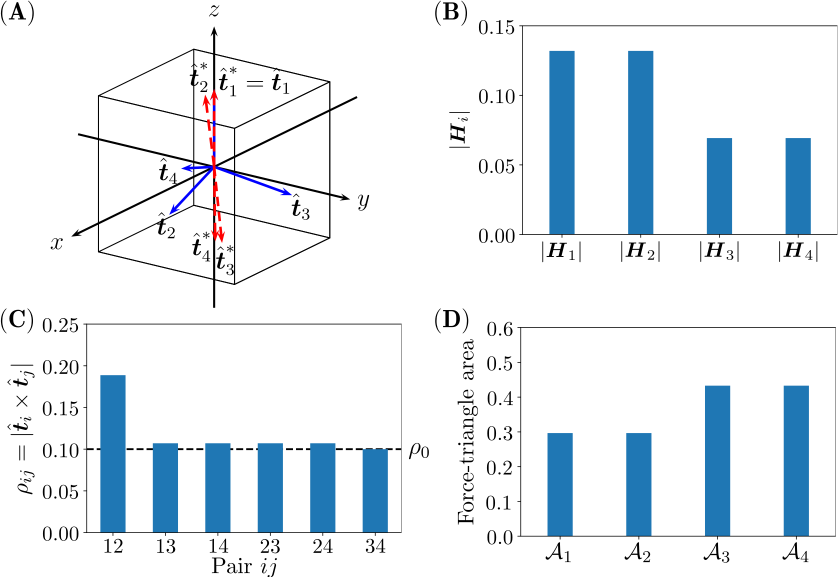}
	\caption{\textbf{Minimum-residual construction for the symmetric non-constructible 4-node defined by Eq.~(\ref{eqnX14}),} obtained by minimizing the dimensionless residual measure $\Phi$ in Eq.~(\ref{eqnX18}). (\textbf{A}) Initial directions $\hat{\bm{t}}_{i}$ and optimized directions $\hat{\bm{t}}_{i}^{*}$. The initial directions correspond to the symmetric non-constructible case in Eq.~(\ref{eqnX14}), rotated into the gauge used in Eq.~(\ref{eqnX20}). (\textbf{B}) Residual magnitudes $|\bm{H}_{i}|$, indicating how the remaining incompatibility is distributed among the four 3-lines. (\textbf{C}) Pairwise separation measures $\rho_{ij}=|\hat{\bm{t}}_{i}^{*}\times\hat{\bm{t}}_{j}^{*}|$, where small values indicate approach to a degenerate local geometry. (\textbf{D}) Areas $\mathcal{A}_{i}$ of the four 3-line force triangles associated with the tetrahedral dual. The minimization is constrained by $\rho_{ij}\ge\rho_{0}$. A constructible 4-node would allow $\Phi_{min}=0$, whereas the non-constructible case shown here retains finite residuals and tends toward a degenerate geometry.}
	\label{fig14}
\end{figure}

The areas $\mathcal{A}_{i}$ of the four force triangles can be evaluated by Eq.~(\ref{eqnX24}). Unlike $|\bm{H}_{i}|$ and $\rho_{ij}$, these areas are independent of the optimized directions and characterize the four individual 3-line force constructions associated with the tetrahedral dual. The face areas $\mathcal{A}_{i}$ provide a useful geometric indicator of which 3-line construction may be most strongly implicated in a non-foldable tetrahedral dual. If one area violates the inequality in Eq.~(\ref{eqn46}), the corresponding face cannot participate in a closed area-vector construction and the associated 3-line is a natural candidate for involvement in a wetting-like elimination. If all four area inequalities are satisfied while $\Delta_{\text{CM}}$ remains negative, the incompatibility is more subtle and must involve the relative arrangement of the six edge lengths, rather than the face areas alone. In that case, the minimum-residual construction provides a complementary measure of how the incompatibility is distributed among the four 3-lines. For the symmetric case
\begin{equation}
	\gamma_{12}=1.9, \quad \gamma_{13}=\gamma_{14}=\gamma_{23}=\gamma_{24}=\gamma_{34}=1 ,
	\label{eqnX14}
\end{equation}
constrained minimization gives $\Phi_{\text{min}}=9.18\times 10^{-4}$. The minimum-residual construction is illustrated in Fig.~\ref{fig14}A. The largest residuals, cf. Fig.~\ref{fig14}B, occur on 3-lines 1 and 2, with $|\bm{H}_{1}|\approx|\bm{H}_{2}|\approx 0.132$, while $|\bm{H}_{3}|\approx|\bm{H}_{4}|\approx 0.069$. Fig.~\ref{fig14}C shows that several pairwise separation measures reach the imposed lower bound $\rho_{0}$, showing that the construction tends toward a degenerate geometry.

\begin{figure}
	\centering
	\includegraphics[width=0.7\textwidth]{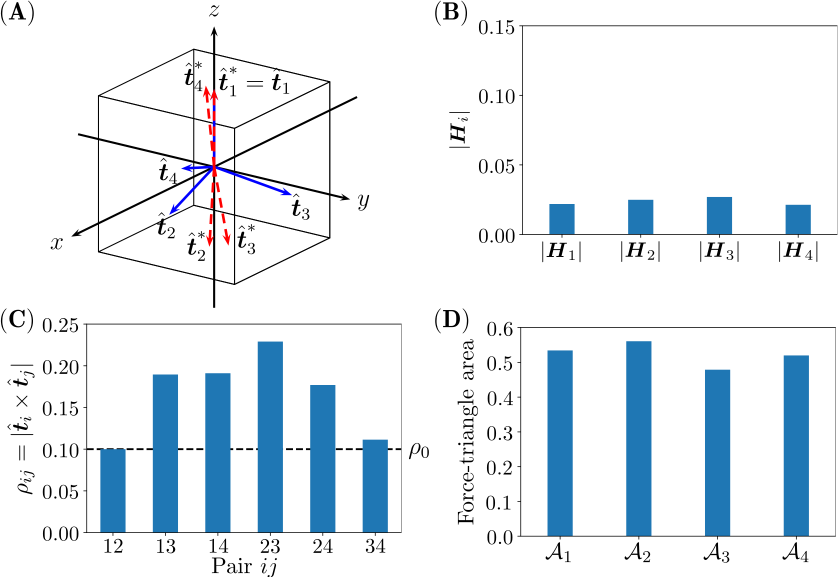}
	\caption{\textbf{Minimum-residual construction for the asymmetric non-constructible 4-node defined by Eq.~(\ref{eqnX22}).} Panels are as in Fig.~\ref{fig14}. The asymmetric energy set breaks the symmetry of the example in Eq.~(\ref{eqnX14}), allowing the residual distribution and pairwise separation measures $\rho_{ij}$ to indicate which 3-line construction carries the largest remaining incompatibility.}
	\label{fig15}
\end{figure}

As an additional illustration, a second case is considered in which the interface energies are obtained by random sampling, giving
\begin{equation}
	\begin{array}{lll}
		\vspace{2mm}
		\displaystyle \gamma_{12} = 0.813231, & \gamma_{13} = 1.330622, & \gamma_{14} = 1.445838, \\
		\displaystyle \gamma_{23} = 1.731364, & \gamma_{24} = 1.398789, & \gamma_{34} = 0.760075
	\end{array}
	\label{eqnX22}
\end{equation}
For this energy set, all four face triangles are constructible while $\Delta_{\text{CM}}=-1.2073$, so the 4-node is non-constructible despite constructibility of all four constituent 3-lines. The resulting minimum-residual construction is obtained at $\Phi_{\text{min}}=4.06\times 10^{-5}$ and is shown in Fig.~\ref{fig15}A. In this asymmetric case, the residuals $|\bm{H}_{i}|$ shown in Fig.~\ref{fig15}B exhibit greater variation but also consistently lower magnitudes than the corresponding results in the symmetric case, shown in Fig.~\ref{fig14}B. The pairwise separations, shown in Fig.~\ref{fig15}C, also vary more. Pair $ij=12$ has reached the constraint $\rho_{0}$, indicating incipient degeneracy. Fig.~\ref{fig14}D and Fig.~\ref{fig15}D show the force-triangle face areas $\mathcal{A}_{i}$. These areas enter the necessary area-vector closure conditions in Eq.~(\ref{eqn46}) for tetrahedral foldability, and provide a geometric indicator of which 3-line construction is most strongly implicated in non-constructibility. In both examples, all four area inequalities are satisfied, showing that the non-constructibility is not detected by the area conditions alone and requires the full Cayley--Menger criterion.

The minimum-residual construction should not be interpreted as a kinetic model. Rather, it identifies the best attainable local arrangement of triple-line directions when the tetrahedral dual is non-constructible and a non-degenerate geometry is enforced. The residual magnitudes $|\bm{H}_{i}|$ indicate how the irreducible incompatibility is distributed among the four 3-lines, while the separation measures $\rho_{ij}$ show whether the construction tends toward collapse of particular interface patches. In this sense, the construction provides a scale-invariant way to characterize non-constructible 4-nodes without introducing arbitrary cutout lengths.

The approach of one or more $\rho_{ij}$ to zero therefore identifies interface patches that tend toward local collapse in the minimum-residual construction and may indicate precursors to topological rearrangement. However, because the construction contains neither kinetics nor information about network connectivity away from the node, it does not determine whether such a collapse produces a 4-line or any other specific subsequent topology.

\subsubsection*{Graphical representation of the fully anisotropic 4-node equilibrium conditions}\label{sect_torque_tetrahedron}
\begin{figure}
	\centering
	\includegraphics[width=0.7\textwidth]{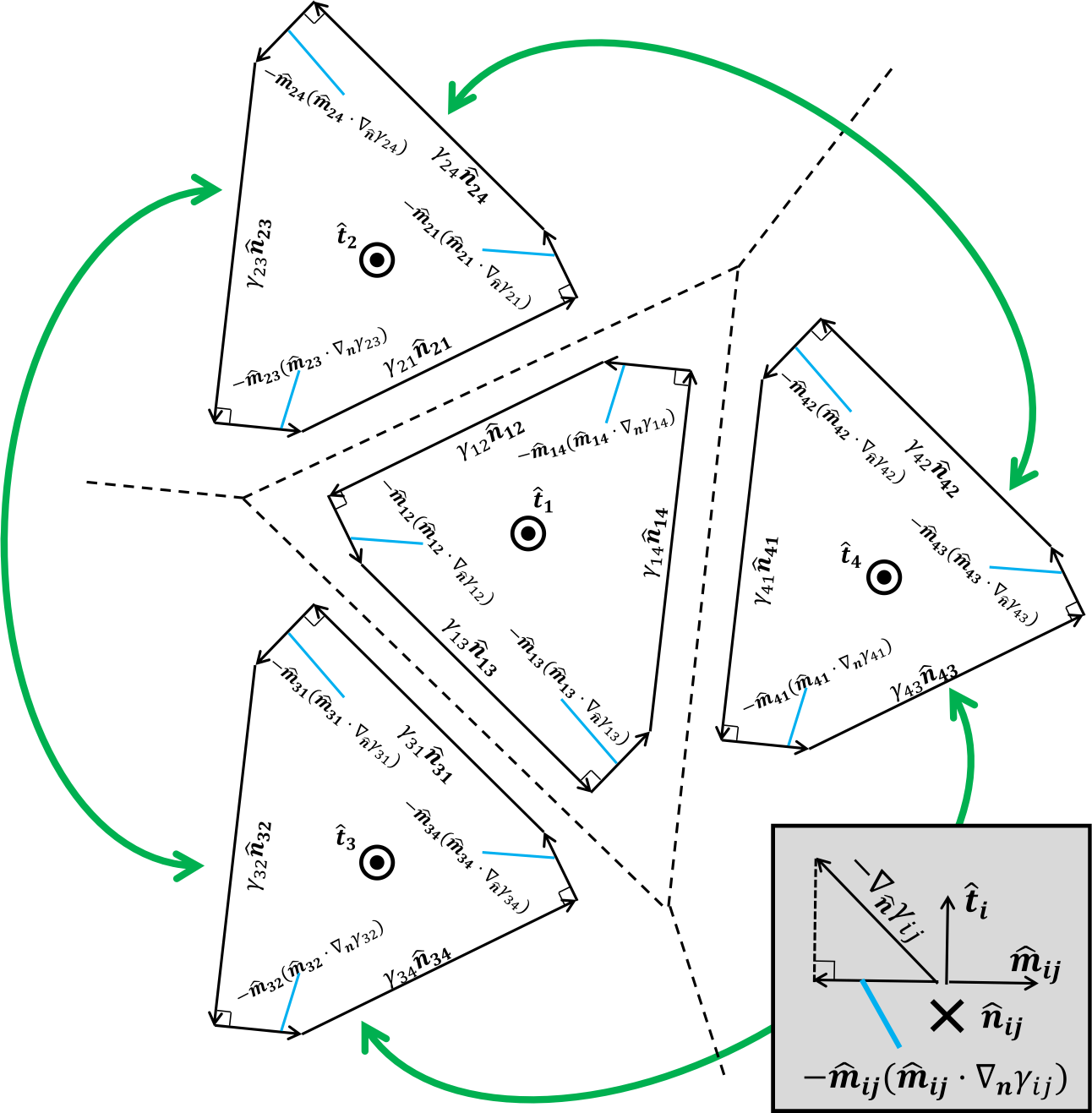}
	\caption{\textbf{Graphical representation of the fully anisotropic 4-node equilibrium conditions.} Unlike the semi-isotropic triangular faces, these hexagons are vector polygons representing force balance; they are not, in general, faces of a single foldable polyhedron. Each polygon corresponds to one 3-line $i$ and represents the rotated Herring condition in Eq.~(\ref{eqn_S58}) in the plane normal to $\hat{\bm{t}}_{i}$. In each polygon, the three capillary segments are the scalar contributions $\gamma_{ij}\hat{\bm{n}}_{ij}$, while the three orthogonal segments are the inclination/torque contributions $-\hat{\bm{ m}}_{ij}(\hat{\bm{m}}_{ij}\cdot\nabla_{\hat{\bm{n}}}\gamma_{ij})$. Their lengths and directions are schematic. In general the magnitudes differ and the directions depend on the signs of the corresponding gradient projections. Closure of each polygon is equivalent to satisfaction of the Herring condition for that 3-line. Matching antiparallel capillary segments, $\gamma_{ij}\hat{\bm{n}}_{ij}=-\gamma_{ji}\hat{\bm{n}}_{ji}$, identify the pairings that reduce to common tetrahedral edges in the semi-isotropic limit. The inset illustrates the projection of the interfacial-energy gradient onto $\hat{\bm{m}}_{ij}$. The dashed lines indicate that the four polygons generally lie in different three-dimensional reference planes.}
	\label{fig18}
\end{figure}

The dual tetrahedron provides a direct geometric representation of the semi-isotropic problem, but does not incorporate inclination/torque terms. Fig.~\ref{fig18} provides a graphical representation of the corresponding fully anisotropic Herring conditions.

We start by rotating the expression for the Herring condition on each 3-line by 90 degrees about that 3-line:
\begin{equation}
	\bm{t}_i\times\bm{H}_{i} = \sum_{j\ne i} \left[\gamma_{ij}\hat{\bm{n}}_{ij} -
	\hat{\bm{m}}_{ij}\left(\hat{\bm{m}}_{ij}\cdot\nabla_{\hat{\bm{n}}}\gamma_{ij}\right)\right]=0.
	\label{eqn_S58}
\end{equation}
The vectors $\gamma_{ij}\hat{\bm{n}}_{ij}$ in Eq.~(\ref{eqn_S58}) are the $90^{\circ}$-rotated representations of the physical capillary-force contributions $\gamma_{ij}\hat{\bm{m}}_{ij}$. The rotation is introduced to expose their pairwise matching in the dual construction. This is the same rotation that is implicitly performed in the construction of the original dual tetrahedron. Besides its natural alignment with the dual construction, this rotation has two distinct advantages. First, it aligns the scalar energy terms explicitly as equal and opposite vectors $\gamma_{ij}\bm{\hat{n}_{ij}}=-\gamma_{ji}\bm{\hat{n}_{ji}}$, and the matching of these antiparallel edges is what produces the tetrahedron from the folding of the four constructions based on each 3-line. Second, it provides an intuitive visualization of the inclination/torque terms as projections of the gradients onto the $\bm{\hat{m}_{ij}}$ (albeit with a minus sign). With the adopted sign convention, we have $\hat{\bm{n}}_{ji}=-\hat{\bm{n}}_{ij}$ and $\nabla_{\hat{\bm{n}}}\gamma_{ji}=-\nabla_{\hat{\bm{n}}}\gamma_{ij}$.

The four hexagons in the unfolded tetrahedron in Fig.~\ref{fig18} each represent Eq.~\ref{eqn_S58} for one of the four values of $i$. The closing of the hexagon, such that it returns precisely to its starting point, is a graphical representation of the Herring condition. These hexagons are irregular, including at least three right angles each if all inclination/torque terms are nonzero. They can also be self-intersecting, depending on the signs of the dot products yielding the inclination/torque segments. The inclination/torque segments in Fig.~\ref{fig18} are drawn schematically. Their magnitudes need not be equal and their directions reverse with the sign of $\hat{\bm{m}}_{ij}\cdot\nabla_{\hat{\bm{n}}}\gamma_{ij}$. The dashed lines in the figure indicate that each such hexagon properly lies in its own reference frame in three dimensions.

After pairwise identification of the antiparallel capillary segments, each would-be tetrahedral vertex is bordered by three inclination/torque segments inherited from three different 3-line polygons. Although all segments within any one polygon are coplanar, lying in the plane normal to its $\hat{\bm{t}}_{i}$, the three torque segments associated with a common would-be tetrahedral vertex generally belong to three different such planes. They therefore need neither be coplanar nor close to form a triangle. Consequently, the four hexagons cannot in general be folded directly into a single polyhedron while preserving all segment identifications. More elaborate non-planar constructions can be devised by introducing additional segments, but they lose the simple geometric interpretation of the semi-isotropic tetrahedron and are not pursued here.

The representation in Fig.~\ref{fig18} also suggests a natural numerical formulation of the fully anisotropic constructibility problem. The four 3-line unit vectors $\hat{\bm{t}}_{i}$ may be used as the independent variables. For prescribed interfacial-energy functions $\gamma_{ij}(\hat{\bm{n}}_{ij})$, the interface normals, capillarity contributions and Herring residuals then follow directly. Constructibility can therefore be tested by seeking a non-degenerate set of directions for which all four residuals vanish simultaneously. The assignment of $\hat{\bm{t}}_i$ in Fig.~\ref{fig18} uses one arbitrary chirality, but the opposite choice is equivalent.




\clearpage

\end{document}